\documentclass[preprint,12pt,authoryear]{elsarticle}

\usepackage{amssymb}
\usepackage{amsmath}
\usepackage{symbols}
\usepackage{times}
\usepackage{graphicx}
\usepackage{color}
\usepackage{soul,url}
\usepackage{longtable}
\usepackage{epsfig}
\usepackage[normalem]{ulem}

\usepackage{lineno}
\usepackage{xspace}
\usepackage{pdfwidgets}
\usepackage{enumerate}
\usepackage{soul}
\usepackage{caption}

\journal{Journal of High Energy Astrophysics}

\def\fermi{Fermi\,J1544-0649}
\def\gr{$\gamma$-ray}
\def\grs{$\gamma$-rays}

\begin{document}
\begin{frontmatter}



\title{Multi-wavelength observations of the BL Lac object Fermi\,J1544-0649: one year after its awakening\tnoteref{label1}}
\tnotetext[label1]{MWL observations of Fermi\,J1544-0649}

\author[1]{P.~H.~T.~Tam\corref{cor1}}
\ead{tanbxuan@sysu.edu.cn}
\cortext[cor1]{Corresponding author}
\author[1]{P.~S.~Pal\corref{cor3}} 
\ead{parthasarathi.pal@gmail.com}
\author[1]{Y.~D.~Cui} 
\author[2]{N.~Jiang}
\author[3]{Y.~Sotnikova} 
\author[2,4]{C.~W.~Yang}
\author[5]{L.~Z.~Wang}
\author[1]{B.~T.~Tang}
\author[6,7,8]{Y.~B.~Li}
\author[6,7,9]{J.~Mao\corref{cor2}}
\ead{jirongmao@mail.ynao.ac.cn}
\author[10]{A.~K.~H.~Kong}
\author[2]{Z.~H.~Zhong}
\author[11]{J. Ding}
\author[12,13]{T.~Mufakharov}
\author[14]{J.~F. Fan}
\author[14]{L.~M.~Dou} 
\author[1]{R.~F.~Shen}
\author[1]{Y.~L.~Ai}
\address[1]{School of Physics and Astronomy, Sun Yat-sen University,\\ 
Guangzhou 510275, People's Republic of China}
\address[2]{CAS Key Laboratory for Researches in Galaxies and Cosmology,\\ 
University of Sciences and Technology of China,\\ Hefei, Anhui 230026, People's Republic of China}
\address[3]{Special Astrophysical Observatory, Russian Academy of Sciences,\\ 
369167, Nizhnij Arkhyz, Russian Federation.}
\address[4]{Polar Research Institute of China,\\ 
451 Jinqiao Road, Shanghai 200136, People's Republic of China.}
\address[5]{Chinese Academy of Sciences South America Center for Astronomy,
\\ China-Chile Joint Center for Astronomy, Camino El Observatorio\\ \#1515, Las Condes, Santiago, Republic of Chile.}
\address[6]{Yunnan Observatories, Chinese Academy of Sciences,\\ Kunming 650011, People's Republic of China.}
\address[7]{Center for Astronomical Mega-Science, Chinese Academy of Sciences,\\ 
20A Datun Road, Chaoyang District, Beijing 100012, People's Republic of China}. 
\address[8]{University of Chinese Academy of Sciences,\\ Beijing 100049, People's Republic of China.}
\address[9]{Key Laboratory for the Structure and Evolution of Celestial Objects,\\ 
Chinese Academy of Sciences, Kunming 650011, People's Republic of China.}
\address[10]{Institute of Astronomy, National Tsing Hua University\\ 
101, Section 2. Kuang-Fu Road, Hsinchu, 30013, Taiwan, R.O.C.}
\address[11]{Department of Astronomy \& Astrophysics, University of California Santa Cruz,\\ 
1156 High Street, Santa Cruz, CA 95060, USA.} 
\address[12]{Shanghai Astronomical Observatory, Chinese Academy of Sciences,\\ 
Shanghai 200030, People's Republic of China.}
\address[13]{Kazan Federal University,\\ 18 Kremlyovskaya St., Kazan 420008, Russian Federation.}
\address[14]{Center for Astrophysics, Guangzhou University,\\
 Guangzhou 510006, People's Republic of China.}

\begin{abstract}
We report observations of a transient source \fermi\ from radio to \grs. \fermi\ was 
discovered by the {\it Fermi-LAT} in May 2017. Follow-up {\it Swift-XRT} observations revealed three flaring episodes 
through March 2018, and the peak X-ray flux is about $10^3$ higher than the {\it ROSAT all-sky survey (RASS)} flux 
upper limit. Optical spectral measurements taken by the {\it Magellan 6.5-m telescope} and the {\it Lick-Shane telescope} 
both show a largely featureless spectrum, strengthening the BL Lac interpretation first proposed by \citet{Bruni18}. 
The optical and mid-infrared (MIR) emission goes to a higher state in 2018, when the flux 
in high energies goes down to a lower level. Our {\it RATAN-600m} measurements at 4.8~GHz and 8.2~GHz do not indicate any
 significant radio flux variation over the monitoring seasons in 2017 and 2018, nor deviate from the archival {\it NVSS} 
flux level. During GeV flaring times, the spectrum is very hard ($\Gamma_\gamma\sim$1.7) in the GeV 
band and at times also very hard (($\Gamma_{\rm X}\lesssim2$) in the X-rays, similar to a high-synchrotron-peak 
(or even an extreme) BL Lac object, making \fermi\ a good target for ground-based {\it Cherenkov telescopes}.

\end{abstract}

\begin{graphicalabstract}
We report observations of a transient source \fermi\ from radio to \grs. \fermi\ was 
discovered by the {\it Fermi-LAT} in May 2017. Follow-up {\it Swift-XRT} observations revealed three flaring episodes 
through March 2018, and the peak X-ray flux is about $10^3$ higher than the {\it ROSAT all-sky survey (RASS)} flux 
upper limit. Optical spectral measurements taken by the {\it Magellan 6.5-m telescope} and the {\it Lick-Shane telescope} 
both show a largely featureless spectrum, strengthening the BL Lac interpretation first proposed by \citet{Bruni18}. 
The optical and mid-infrared (MIR) emission goes to a higher state in 2018, when the flux 
in high energies goes down to a lower level. Our {\it RATAN-600m} measurements at 4.8~GHz and 8.2~GHz do not indicate any
 significant radio flux variation over the monitoring seasons in 2017 and 2018, nor deviate from the archival {\it NVSS} 
flux level. During GeV flaring times, the spectrum is very hard ($\Gamma_\gamma\sim$1.7) in the GeV 
band and at times also very hard (($\Gamma_{\rm X}\lesssim2$) in the X-rays, similar to a high-synchrotron-peak 
(or even an extreme) BL Lac object, making \fermi\ a good target for ground-based {\it Cherenkov telescopes}.

\\
\includegraphics[angle=270,scale=0.6]{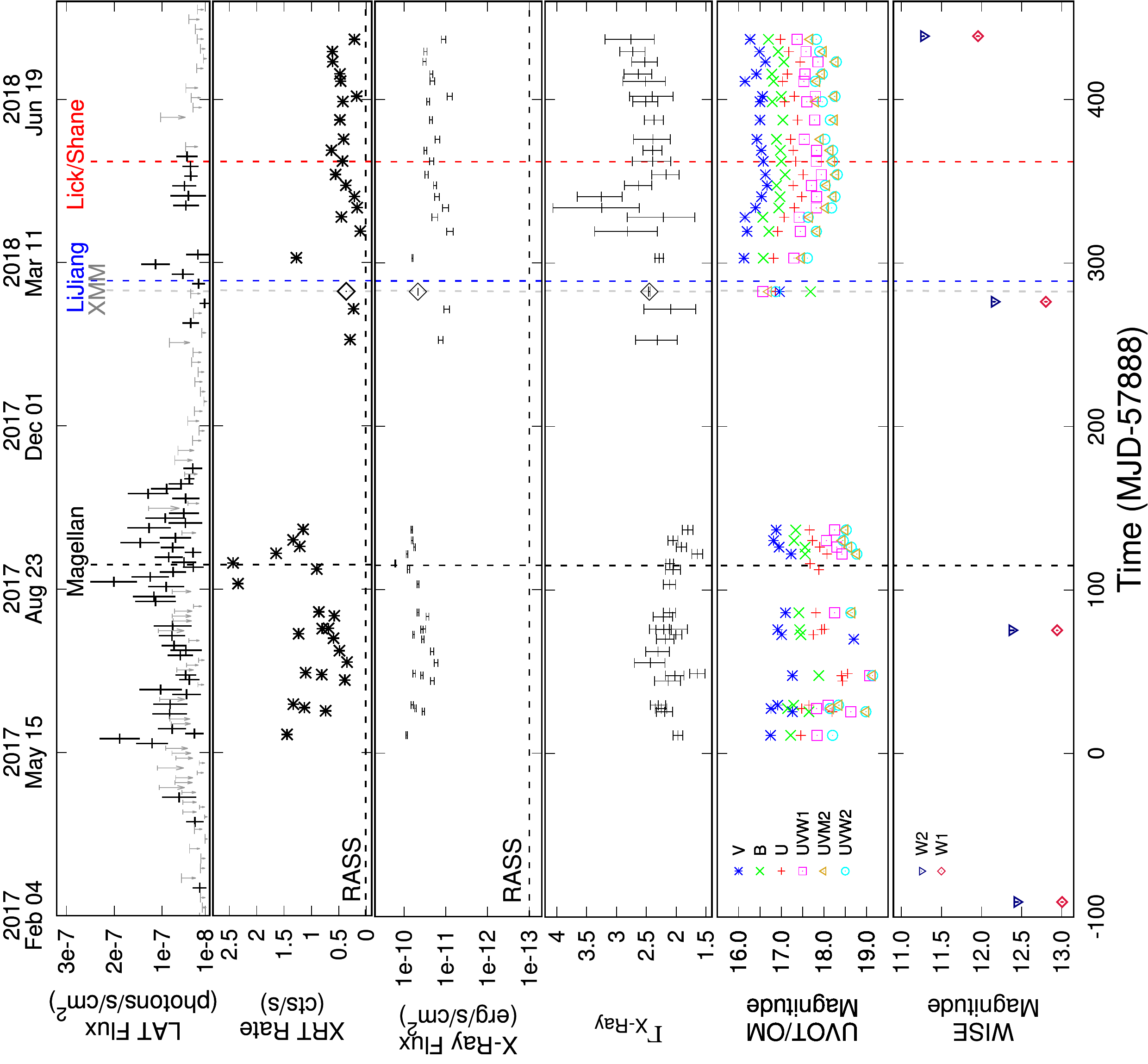}
\end{graphicalabstract}


\end{frontmatter}


\section{Introduction}
\label{sec:into}

Super-massive black holes (SMBHs; with mass $\ga$10$^6\,M_{\sun}$) locate at the centres of galaxies. 
By accreting a large enough amount of mainly gaseous materials, a SMBH can generate 
luminous emission over the whole electromagnetic spectrum, referred as an Active Galactic 
Nuclei~\citep[AGN;][]{Lynden-Bell_69}. Sometimes, an AGN could generate powerful jets, when 
the jet direction nearly co-aligns with the line of sight, this AGN is seen as a blazar 
from the Earth - characterized by large variability at all wavelengths 
and usually accompanied by \gr\ emission \citep[e.g.,][]{Urry95}. Variability at all wavelengths is a 
defining feature of blazars, so some blazars may be seen as a transient. They may remain quiet and only become 
bright in a relatively short time scale (e.g., months), and all-sky high-energy monitors like {\it Fermi-LAT} 
and {\it MAXI} may catch such rare events.

An optical transient, ASASSN-17gs, or AT2017egv, was detected at V=17.3 mag on 2017-05-25 09:36 UT 
(i.e., contemporaneous to the {\it LAT} transient detections). The host galaxy, 2MASX~J15441967-0649156, was 
found to be at a spectroscopic Redshift of z=0.171, using the {\it MDM 2.4m Hiltner telescope} on the night of 
2017 June 14 UT \citep{Chornock17}. 

The persistent radio source at the same position, NVSS~J154419-064913, has a flux density of 46.6~mJy 
at 1.4~GHz in 1996/1997 \citep{NVSS_survey}. This flux density, at a Redshift of 0.171, corresponds to 
4$\times$10$^{31}$~erg~s$^{-1}$~Hz$^{-1}$, a radio luminosity which is above most of known radio-loud AGN 
\citep[see, e.g., Fig.~11 of][]{2014ARA&A..52..589H}. {\it GMRT} observed 66.6$\pm$8.4 mJy at 150 MHz between April 2010 
and March 2012 \citep{GMRT_survey}. In this work, we present detailed data analysis in \grs\ and X-rays, 
optical photometry and spectroscopy, radio flux monitoring in following sections. We further discuss our 
main findings in Section.~\ref{sec:discuss}, including the characteristic blazar SED peaked at X/\gr,
 the fast X-ray variation with a time scale down to 1 hour, the mysterious continuum optical component with 
week-scale variations. 

\section{Observed Evolution of the High-Energy Emission}
\label{sec:high_lcj1544_6-eps-converted-to.pdf}

\subsection{\gr\ Emission}

The {\it LAT} detector is an all-sky monitor at energies from several tens of MeV to more than 300 GeV \citep{lat_technical}. 
The \gr\ data\footnote{provided by the FSSC at \url{http://fermi.gsfc.nasa.gov/ssc/}} used in this work 
were obtained using the {\it Fermi-LAT} between 2008 August 4 and 2018 August 15. We used the \texttt{Fermi Science Tools 
v10r0p5 package} to reduce and analyze the data. Pass 8 data classified as ``source'' events were used. 
To reduce the contamination from Earth albedo \grs, events with zenith angles greater than 
100$\arcdeg$ were excluded. The instrument response functions ``P8R2\_SOURCE\_V6'' were used.

To constrain the normalization of diffuse background and the spectral parameters of nearby sources for latter 
shorter-duration analysis, we first carried out a binned maximum-likelihood analysis (\texttt{gtlike}) of a 
rectangular region of 21$\arcdeg\times$21$\arcdeg$ centered on the position of \fermi, using 9-years of data. 
To this end, we subtracted the background contribution by including the Galactic diffuse model (gll\_iem\_v06.fits) 
and the isotropic background (iso\_P8R2\_SOURCE\_ V6\_v06.txt), as well as the third {\it Fermi-LAT} 
catalog \citep[3FGL;][]{lat_3rd_cat} sources within 25$^\circ$ away from \fermi. 
The recommended spectral model for each source as in the 3FGL~catalog was used, while we modeled a putative 
source at the position of \fermi\ with a power-law (PL):

\begin{equation}
\frac{dN}{dE} = N_0 \left(\frac{E}{E_0}\right)^{-\Gamma},
\end{equation}

where the normalization $N_0$ and spectral index $\Gamma$ were allowed to vary. The normalization parameter 
values for the Galactic and isotropic diffuse components, and sources within 6$\arcdeg$ from \fermi\ were allowed 
to vary as well. Other parameters were held fixed. 

\begin{figure*}
\centering
\includegraphics[width=11cm]{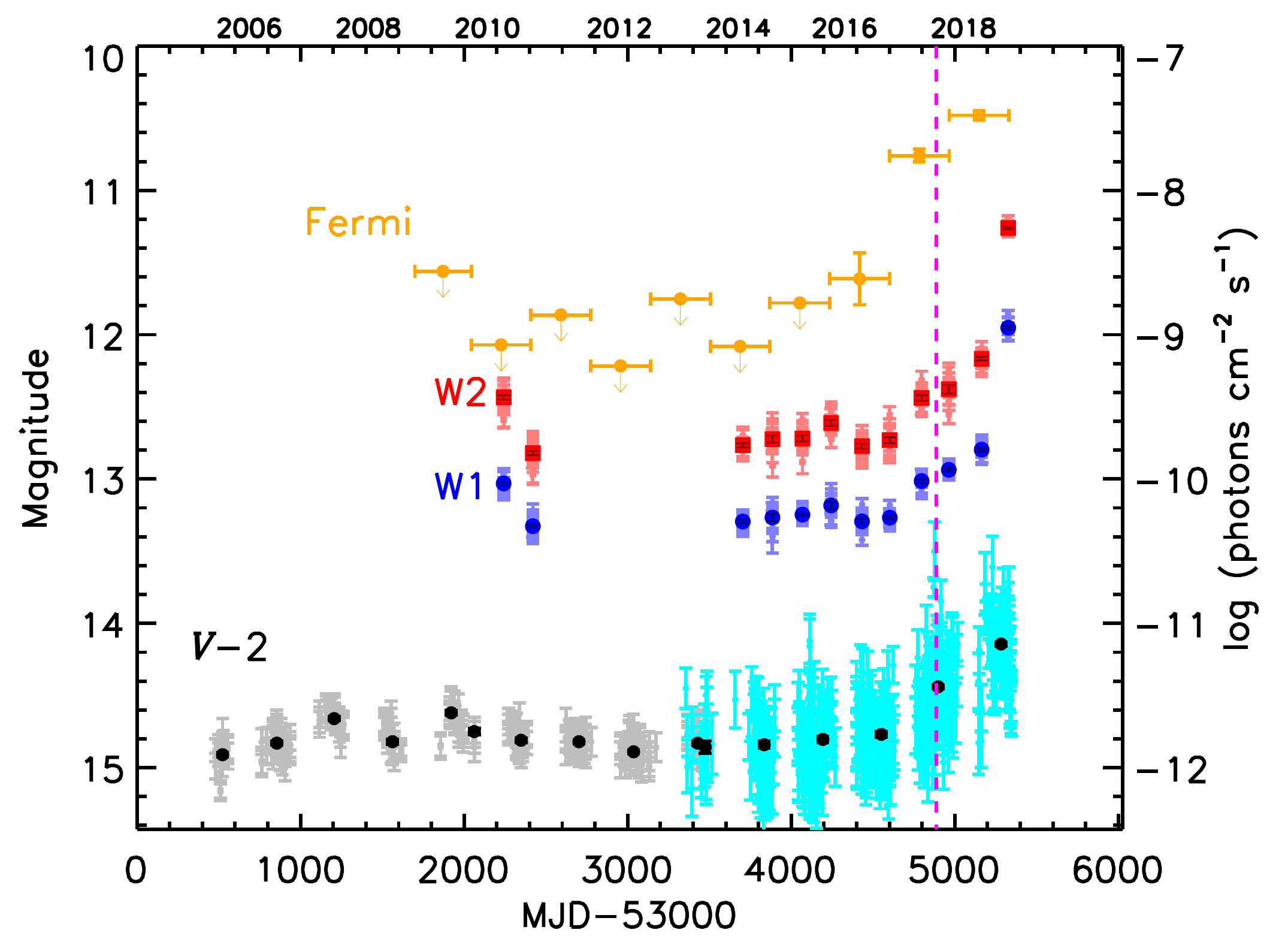}
\caption{The \gr, optical and MIR light curves of \fermi. The {\it Fermi-LAT} \gr\ photon flux (orange) show an increase 
in 2017 and 2018 relative to previous years. The $V$-band data are collected from public releases of {\it CRTS} (grey) 
and {\it ASAS-SN} (cyan); the MIR data are drawn from {\it WISE} database in W1 (blue) and W2 (red). The magenta line 
indicates MJD 57888 (i.e., 2017 May 15), when \fermi\ was discovered and so the {\it Fermi} flux seen in the second last 
data is mostly from thereafter.
}
\label{fig:oirlc}
\end{figure*}

\begin{figure*}
\centering
\includegraphics[angle=270,scale=0.5]{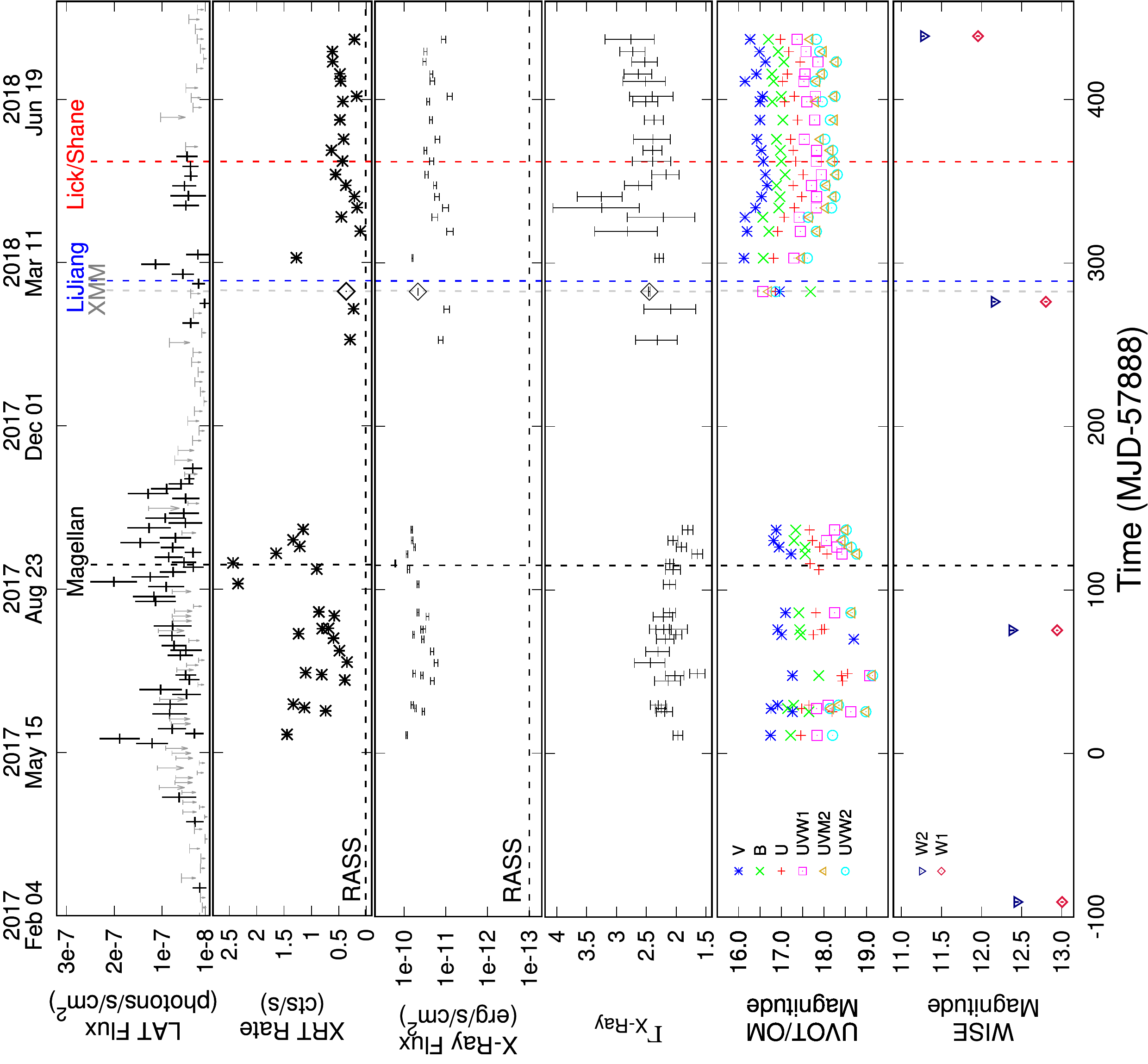}
\caption{{\it From top to bottom}: The {\it Fermi-LAT} 0.1-300 GeV photon flux, {\it Swift-XRT} 0.3-10~keV count rate, 
energy flux, photon index, {\it Swift-UVOT} magnitudes of various filters, and WISE magnitudes of \fermi\ as seen 
between February 2017 and August 2018. The three arrows in the top panel indicates the dates of spectroscopic 
observations. When the source is not detected by {\it Fermi-LAT}, 90\% confidence level upper limits were derived and 
are plotted in grey. The error bars in the second, fifth and sixth panels are smaller than the symbols. 
For comparison, the dashed, horizontal lines show the upper limits (a count rate of 0.01 cts/s and an energy 
flux of $10^{-13}$~erg~cm$^{-2}$~s$^{-1}$) estimated from the {\it RASS} observations. A large increase in X-ray 
flux (up to three orders-of-magnitude) is clearly seen. In the second, third, and fourth panel the black 
diamonds represent the {\it XMM-EPIC PN} result from the 2018 February 21 observation. In the second panel the 
shown ({\it SWIFT-XRT} equivalent) count rate is converted from {\it XMM-EPIC PN} count rate with the help of {\it WebPIMMS}.
\label{fig:mwl_lc}} 
\end{figure*}

Using the whole data set from the first 8.6 years, we did not detect any source at the \fermi\ position. 
\gr\ flux over monthly time bins were also deduced by letting the normalization and photon index to 
vary in the iteration. No significant detection (i.e., above TS=12) was found until May 2017. The same was 
done for year time scale, and only during the last two years was the source detected significantly 
(see Fig.~\ref{fig:oirlc}). We thus confirm this transient nature as a recent event. With the 8.6-year 
background model at hand, we carried out maximum likelihood analysis on 3-day/6-day bins from April 2017 
to August 2018, and the results are plotted in Fig.~\ref{fig:mwl_lc}. The average \gr\ photon index during 
the {\it Fermi} flares is about 1.7. Flux upper limits were deduced and plotted whenever TS$<$9. It can be seen 
that the flaring period lasts for 180 days since MJD 57888, and is composed of two major flares at May to 
June and August to September 2017. There is a third major flare, though smaller in magnitude, in March 2018. 
All three major \gr\ flares are accompanied by X-ray flare seen by {\it Swift-XRT}.
The \gr\ flux goes to a lower level of activity in 2018, as compared to May through October in 2017.

\subsection{X-ray Emission}
\subsubsection{{\it XMM-Newton} observation}
\begin{figure*}
\includegraphics[angle=270,scale=0.255]{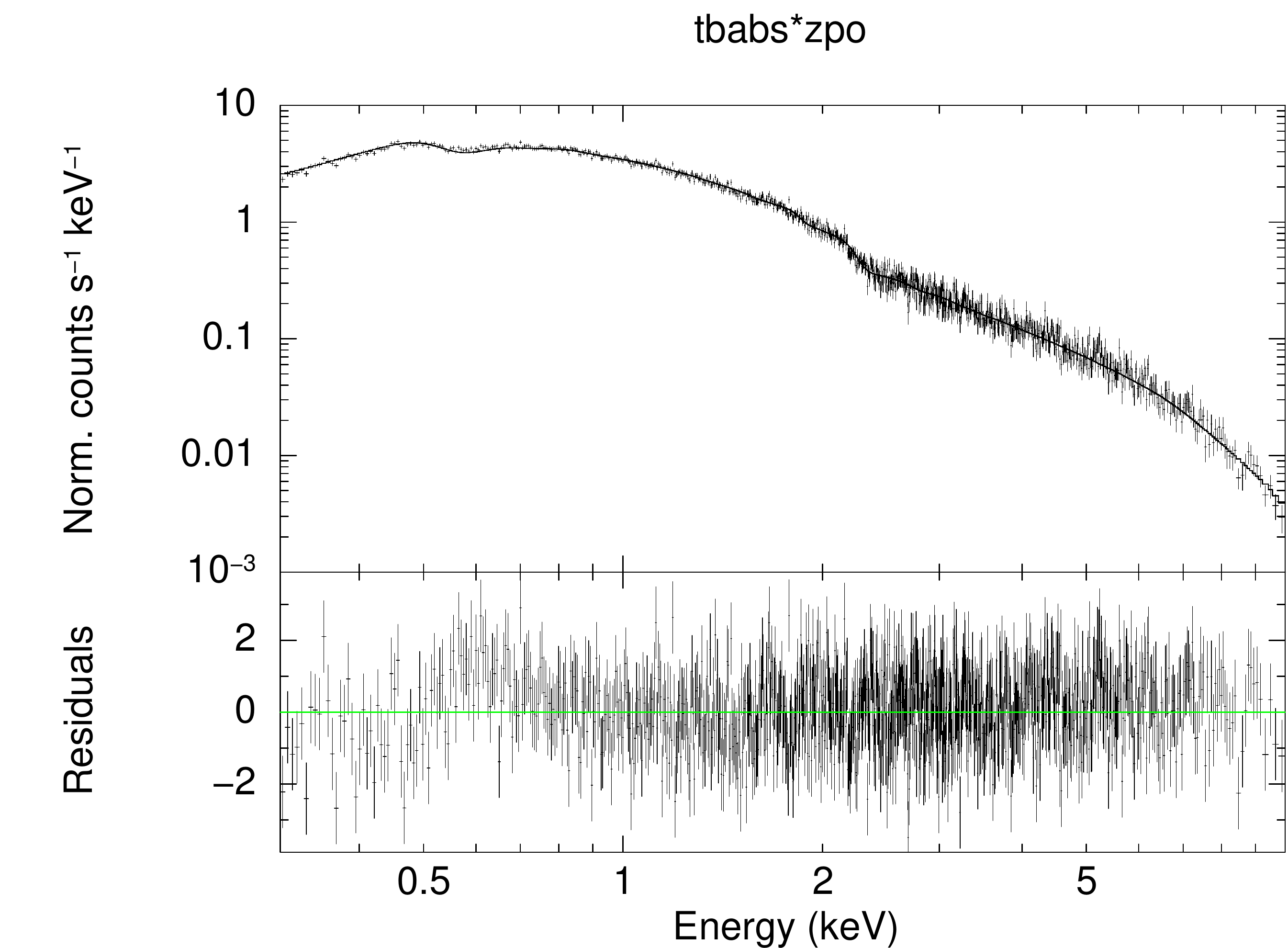}
\includegraphics[angle=270,scale=0.255]{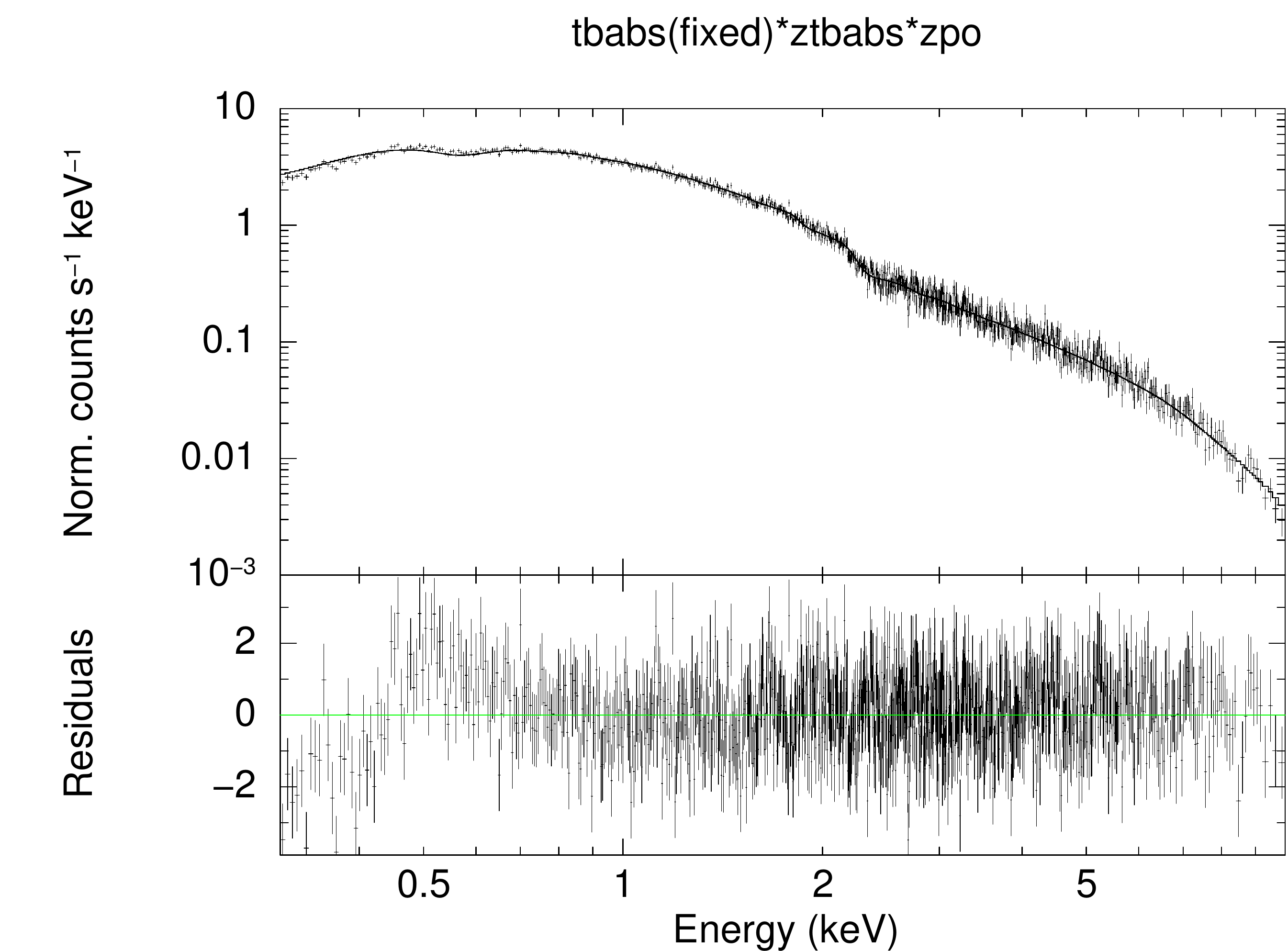}
\includegraphics[angle=270,scale=0.255]{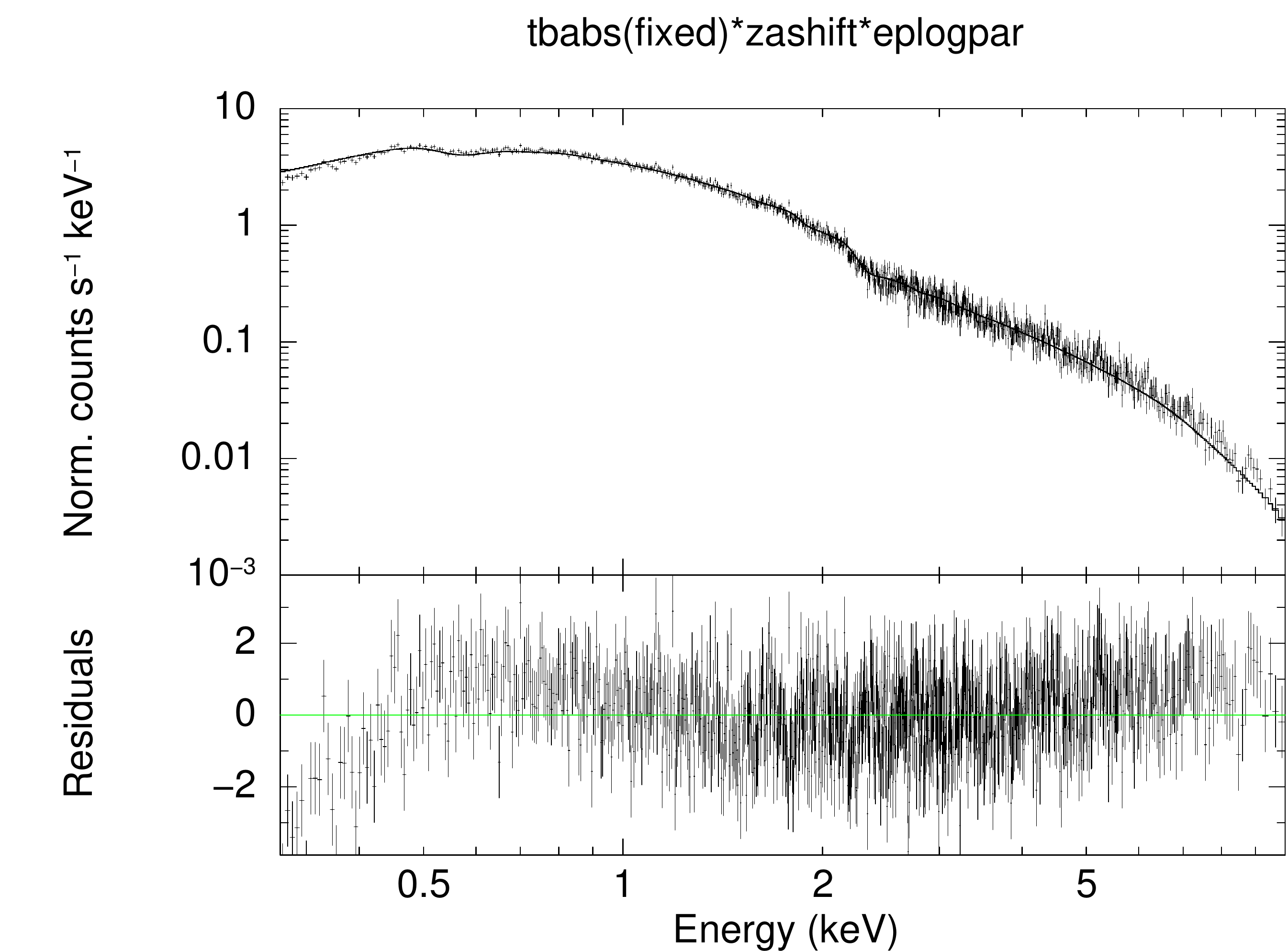}
\caption{{\it EPIC-pn} spectrum fitted with different models for the whole observation
\label{pn_spec}}
\end{figure*}

{\it XMM-Newton} DDT observation of \fermi\ (obs-id: 0811213301) was performed on 
21st February, 2018 (MJD 58170) for about 58 ksec. We used {\it EPIC-PN} data for the X-ray analysis, 
as they have higher sensitivity than {\it EPIC-MOS} data. We verify that the MOS data return consistent 
results as the pn data. The data reduction was performed with the software SAS (version 16.1), using 
the most updated calibration files (updated on May 2018). The event files were processed using `epproc' 
with `bad' (e.g., `hot', `dead', `flickering') pixels removed. 
The periods with high background events were examined and excluded by inspecting the light curves in 
the energy band 10-12 keV. As the X-rays of \fermi\ is bright and the pile-up effect 
is apparent in the source center, we extracted the source events from an annular region 
with inner radius of 7.5$\arcsec$ and outer radius of 40$\arcsec$, using single and double events 
(PATTERN$\leq$4, FLAG=0). The background events were collected from a source-free circular region of 
radius 40$\arcsec$ within net exposure time of 29.93 ks, and are composed of a total of 165 thousand net source 
counts in 0.3--10 keV band.  
We grouped the pn spectra to have at least 25 counts in each bin, and we adopt the $\chi^2$ statistic 
for the spectral fits. The fitted pn spectra are shown in Fig.~\ref{pn_spec}.
The spectral analysis were performed using {\it XSPEC(v12.9.1m)}. The uncertainties are given at 90\% confidence 
levels for one parameter.

At first, a simple neutral-hydrogen absorbed power-law (PL) model (tbabs $\times$ zpo) was used, 
and we obtained $\chi^2/dof$ = 946/900 with $n_{\rm H}=(14.7\pm0.27)\times10^{20}$~cm$^{-2}$. 
To understand the absorption and the spectrum of the object, we compared different models. 
A simple neutral-hydrogen absorbed PL model (tbabs $\times$ zpo), in which the neutral hydrogen 
column density is fixed at the Galactic value of $8.98\times10^{20}$~cm$^{-2}$ \citep{Kalberla05}, is used.
It results in $\chi^2/dof$ = 2397/901, showing that the model does not work for the data. 
We then added an intrinsic absorber ($ztbabs$) into the model, where the Redshift is fixed at 0.17 \citep{Chornock17}. 
The fit is then much improved ($\chi^2/dof$ =1015/900), and results in an intrinsic absorber with neutral hydrogen 
column density of $(7.0\pm2.0)\times10^{20}$~cm$^{-2}$. However, there is some systematics in the lowest 
(0.3--0.8~keV) and highest energy ($>$7~keV).
When trying an ionized absorber model \citep{Zdziarski95}, the ionisation parameter is essentially zero, 
indicating that the absorber is not heavily ionized.

We have also used a log-parabolic (LP) model \citep[$eplogpar$, in which $N(E)=10^{-b(log(E/E_{\rm p}))^2/E^2}$ 
often used for blazars;][]{2007A&A...466..521T}. Here we fix the column density at the Galactic value.
We obtained ($\chi^2/dof$ = 1121/901) for this model with peak energy $E_{\rm p}=(0.85\pm0.03)$~keV and
a curvature $b$ of $0.40\pm0.02$. These results are shown in Table.~\ref{xtab1}. We also tried to allow the 
column density to vary in the LP model, but the fit parameters are not stable. 
Based on the above analysis, the PL model (with intrinsic absorption) and the LP model 
(without intrinsic absorption) can both describe the whole data set well. 

\begin{table*}
\scriptsize{
\addtolength{\tabcolsep}{-3pt}
\begin{center}
\caption{{\it XMM-Newton} EPIC-pn spectral analysis result of \fermi~on 2018 February 21. (I)~PL represents tbabs$\times$zpo, 
(II)~PL represents tbabs$\times$ztbabs$\times$zpo and (III)~LP represents tbabs$\times$zashift$\times$eplogpar model
components. \label{xtab1}}
\begin{tabular}{lccccccc}
\hline
Models & $n_{\rm H}$(Galactic) & $n_{\rm H}$  & $\Gamma$ & $E_{\rm p}$  & $b$ & $10^{11}\times$Flux & $\chi^2_\nu$  \\
 & ($10^{22}$~cm$^{-2}$)  &  ($10^{22}$~cm$^{-2}$)  &    & (keV) &    & (erg~s$^{-1}$~cm$^{-2}$) & (dof)  \\
\hline
(I)~PL & \dots & $0.14\pm0.003$ & $2.47\pm0.01$ & \dots & \dots & $4.68\pm0.02$ & 1.05(900) \\
(II)~PL & 0.0898(fixed) & $0.07\pm0.01$  & $2.45\pm0.01$ & \dots & \dots & $4.57\pm0.02$ & 1.128(900) \\
(III)~LP & 0.0898(fixed) & \dots & \dots & $0.85\pm0.03$ & $0.40\pm0.02$  &  $5.13\pm0.02$ & 1.246(901) \\ 
\hline
\end{tabular}
\end{center}}
\end{table*}

\begin{figure*}
\centering
\includegraphics[width=8cm,angle=270]{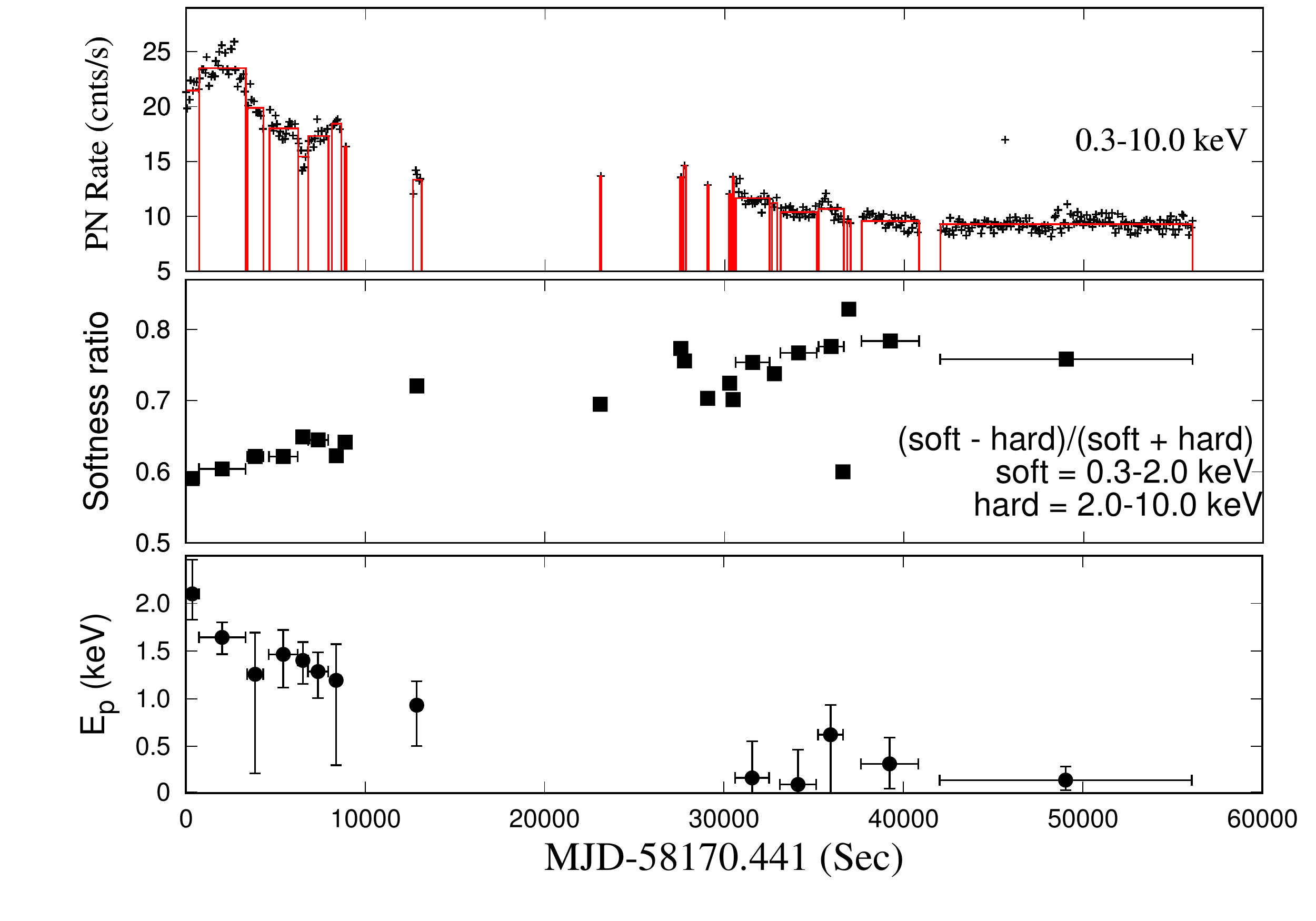}
\caption{{\it XMM-Newton EPIC-PN} cleaned light curve for 0.3--10.0 keV (first panel) energy bands 
with 100~s time bins and the Bayesian block intervals with 95\% statistical 
significance in red color, along with the softness ratio (second panel) and evolution of the 
peak energy ($E_{\rm p}$) in the log-parabola model (third panel).
\label{pn_dyna}}
\end{figure*}

\begin{figure*}
\centering
\includegraphics[width=8cm,angle=270]{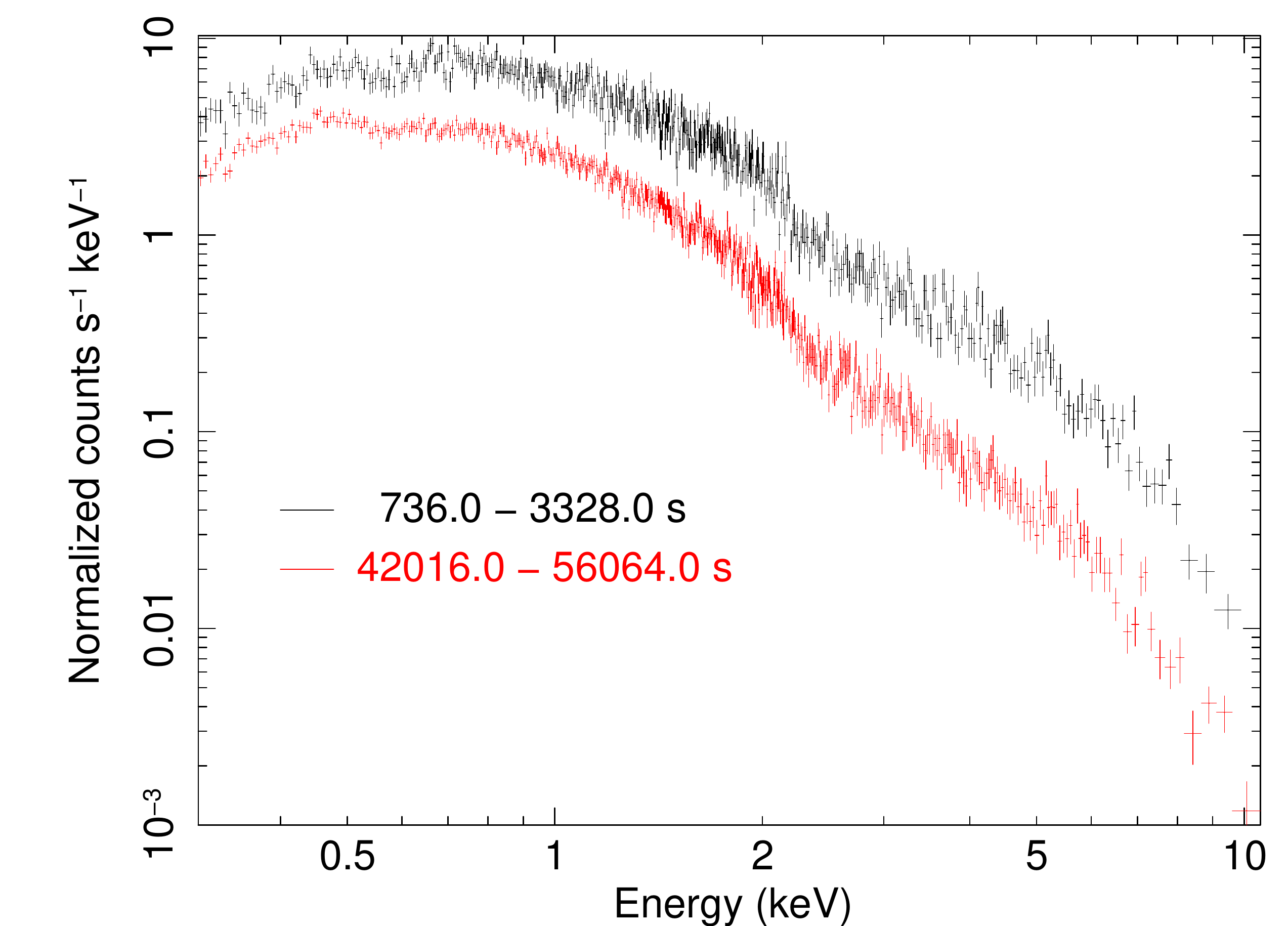}
\caption{{\it XMM-Newton EPIC-PN} cleaned time-resolved spectra for comparison.
The color codes represent the time intervals (as defined in Table.~\ref{xtab2}) 
as follows: black - 736--3328 sec, red - 42016--56064 sec.
\label{pn_dspec}}
\end{figure*}

We then looked into the timing analysis. 
The {\it EPIC-PN} cleaned light curve in 0.3--10.0 keV band in 100 s bin is shown in Fig.~\ref{pn_dyna}. 
The background was subtracted and the instrumental effect was corrected using the task `epicclorr'. 
The X-ray light curve shows that \fermi\ varies on timescales of a few ks. 
This prompted us to perform time-resolved spectral analysis. We divided the whole observation into 
40 Bayesian blocks \citep{bayes}, calculated with 95\% statistical significance using Python module 
{\it Astropy} \citep{astropy:2013, astropy:2018}. 
In Fig.~\ref{pn_dspec} we show the spectra for the two different block intervals. 
In Fig.~\ref{pn_dyna} top panel blocks are shown with red color along with the cleaned light curve.
In the second panel we have calculated the softness ratio between (0.3--2.0) keV and (2.0--10.0) keV energy bands 
for the Bayesian block intervals. 
For time resolved spectral analysis we ignored the blocks with lesser number of photons. We used 13 block 
intervals for time dependent spectral analysis. 
We employed the PL model (tbabs $\times$ zpo) first. We also used the LP model in the 
time-resolved spectra. The time-resolved spectral analysis results are shown in Table.~\ref{xtab2}. 
It can be seen that the spectrum becomes harder when brighter, i.e., when the count rate decreases, 
the softness ratio increases and the peak energy (in the LP model) decreases (see, third panel Fig.~\ref{pn_dyna}).
This shows that the goodness-of-fit (i.e., reduced $\chi^2$) in the time-integrated 
fits is affected by the changing spectrum. To conclude, the PL model is as good as the LP model based on 
the goodness-of-fit (especially for the short time interval spectral fits).
Based on the goodness-of-fit for time-resolved {\it XMM-Newton} spectra, we found that the LP model and 
the PL model can both describe the data well. However, from the broad-band SED, the X-rays represent the 
synchrotron bump, and it is anticipated that the X-ray spectrum is curved. 

\begin{table*}
\begin{center}
\caption{{\it XMM-Newton EPIC-PN} time-resolved spectral analysis results of \fermi\ on 2018 February 21 \label{xtab2}}
\begin{tabular}{ccccc}
\hline
\multicolumn{5}{c}{Model (I): Power law}\\
\hline
Interval & $n_{\rm H}$ & $\Gamma$ & $10^{11} \times Flux$ & $\chi^2_\nu$ \\
(s) & ($10^{22}$~cm$^{-2}$) & & (erg~s$^{-1}$~cm$^{-2}$) & (dof) \\
\hline
0-736 & $0.14 \pm 0.02$ & $2.15 \pm 0.07$ & $8.12 \pm 0.18$ & 0.85(179) \\ 
736-3328 & $0.13 \pm 0.01$ & $2.16 \pm 0.03$ & $8.55 \pm 0.10$ & 1.17(458) \\ 
3424-4320 & $0.11 \pm 0.02$ & $2.12 \pm 0.07$ & $6.86 \pm 0.14$ & 1.05(196) \\ 
4640-6240 & $0.13 \pm 0.01$ & $2.17 \pm 0.05$ & $6.40 \pm 0.11$ & 0.98(302) \\ 
6240-6816 & $0.17 \pm 0.03$ & $2.40 \pm 0.11$ & $6.06 \pm 0.18$ & 0.96(108) \\ 
6816-7936 & $0.14 \pm 0.02$ & $2.27 \pm 0.07$ & $6.30 \pm 0.13$ & 0.94(208) \\ 
8128-8640 & $0.14 \pm 0.03$ & $2.24 \pm 0.11$ & $6.61 \pm 0.19$ & 1.23(111) \\ 
12640-13120 & $0.16 \pm 0.04$ & $2.51 \pm 0.15$ & $5.10 \pm 0.18$ & 0.94(79) \\ 
30624-32512 & $0.15 \pm 0.04$ & $2.66 \pm 0.10$ & $4.36 \pm 0.10$ & 1.06(166) \\ 
33120-35136 & $0.13 \pm 0.09$ & $2.66 \pm 0.11$ & $3.78 \pm 0.09$ & 0.98(155) \\ 
35232-36640 & $0.21 \pm 0.05$ & $2.90 \pm 0.15$ & $4.91 \pm 0.14$ & 1.34(114) \\ 
37632-40832 & $0.17 \pm 0.04$ & $2.83 \pm 0.09$ & $3.93 \pm 0.08$ & 0.87(210) \\ 
42016-56064 & $0.14 \pm 0.02$ & $2.65 \pm 0.04$ & $3.47 \pm 0.03$ & 1.01(451) \\ 
\hline
\multicolumn{5}{c}{Model (III): Log-parabolic (absorption fixed at the Galactic value)}\\
\hline
Interval & $E_{\rm p}$ & $b$ & $10^{11} \times S_{\rm p}$ & $\chi^2_\nu$ \\
(s) & (keV) & & (erg~s$^{-1}$~cm$^{-2}$) & (dof) \\
\hline
0-736 & $2.10 \pm 0.31$ & $0.32 \pm 0.12$ & $3.04 \pm 0.50$ & 0.86(179) \\ 
736-3328 & $1.64 \pm 0.17$ & $0.24 \pm 0.06$ & $3.26 \pm 0.50$ & 1.15(458) \\ 
3424-4320 & $1.26 \pm 0.74$ & $0.15 \pm 0.12$ & $2.67 \pm 0.50$ & 1.05(196) \\ 
4640-6240 & $1.47 \pm 0.30$ & $0.21 \pm 0.09$ & $2.44 \pm 0.50$ & 0.98(302) \\ 
6240-6816 & $1.40 \pm 0.22$ & $0.53 \pm 0.19$ & $2.16 \pm 0.50$ & 0.97(108) \\ 
6816-7936 & $1.28 \pm 0.24$ & $0.33 \pm 0.12$ & $2.40 \pm 0.50$ & 0.93(208) \\ 
8128-8640 & $1.19 \pm 0.63$ & $0.23 \pm 0.18$ & $2.49 \pm 0.50$ & 1.26(111) \\ 
12640-13120 & $0.93 \pm 0.34$ & $0.49 \pm 0.25$ & $1.93 \pm 0.50$ & 0.93(79) \\ 
35232-36640 & $0.62 \pm 0.47$ & $0.60 \pm 0.30$ & $1.76 \pm 0.51$ & 1.35(114) \\ 
\hline
\end{tabular}
\end{center}
\end{table*}

\subsubsection{{\it Swift-XRT} observations}
Since 2017 May 26, Swift monitoring observations have been performed. Here we present all XRT results obtained until 2018 July 25.
For {\it Swift-XRT} data reduction, the level 2 cleaned event files of {\it SWIFT-XRT} 
were obtained from the events of photon counting (PC) mode data with {\it xrtpipeline}. 
The spectra were extracted from a circular region in the best source position with 20$\arcsec$ radius. 
The background was estimated from an annular region in the same position with radii from 30$\arcsec$ to 60$\arcsec$.
The ancillary response files (arfs) were extracted with {\it xrtmkarf}. 
The PC redistribution matrix file (rmf) version (v.12) was used in the spectral fits. {\it XRT} spectra are grouped with 5 counts per bin.
{\it XRT} spectrum is then analysed with {\it XSPEC(v12.9.1m)} in the similar process as the {\it XMM-Newton EPIC-PN} spectra. 
We here fix the absorption column density to be $n_{\rm H}=(14.7\pm0.27)\times10^{20}$~cm$^{-2}$, the value found 
from the {\it XMM-Newton} analysis.
From the fitted spectra, unabsorbed flux values were calculated from 0.3--10 keV in cgs units for all 
observations. The {\it Swift-XRT} light curve and evolution of the power-law index are plotted in Fig.~\ref{fig:mwl_lc}. 
Some but not all {\it Swift} spectra can be fitted with the LP model (tbabs*zashift*eplogpar). 
After the discovery of the first major flare in 2017 May 26, the {\it Swift} X-ray light curve shows a second major flaring 
episode in August and September 2017. After that, the high energy flux has decreased to a lower level, besides a third 
flaring episode in February to March 2018 (see Fig.~\ref{fig:mwl_lc}).

\subsubsection{X-ray correlation properties}
\label{sect:x_corr}

\begin{figure*}
\includegraphics[width=\textwidth]{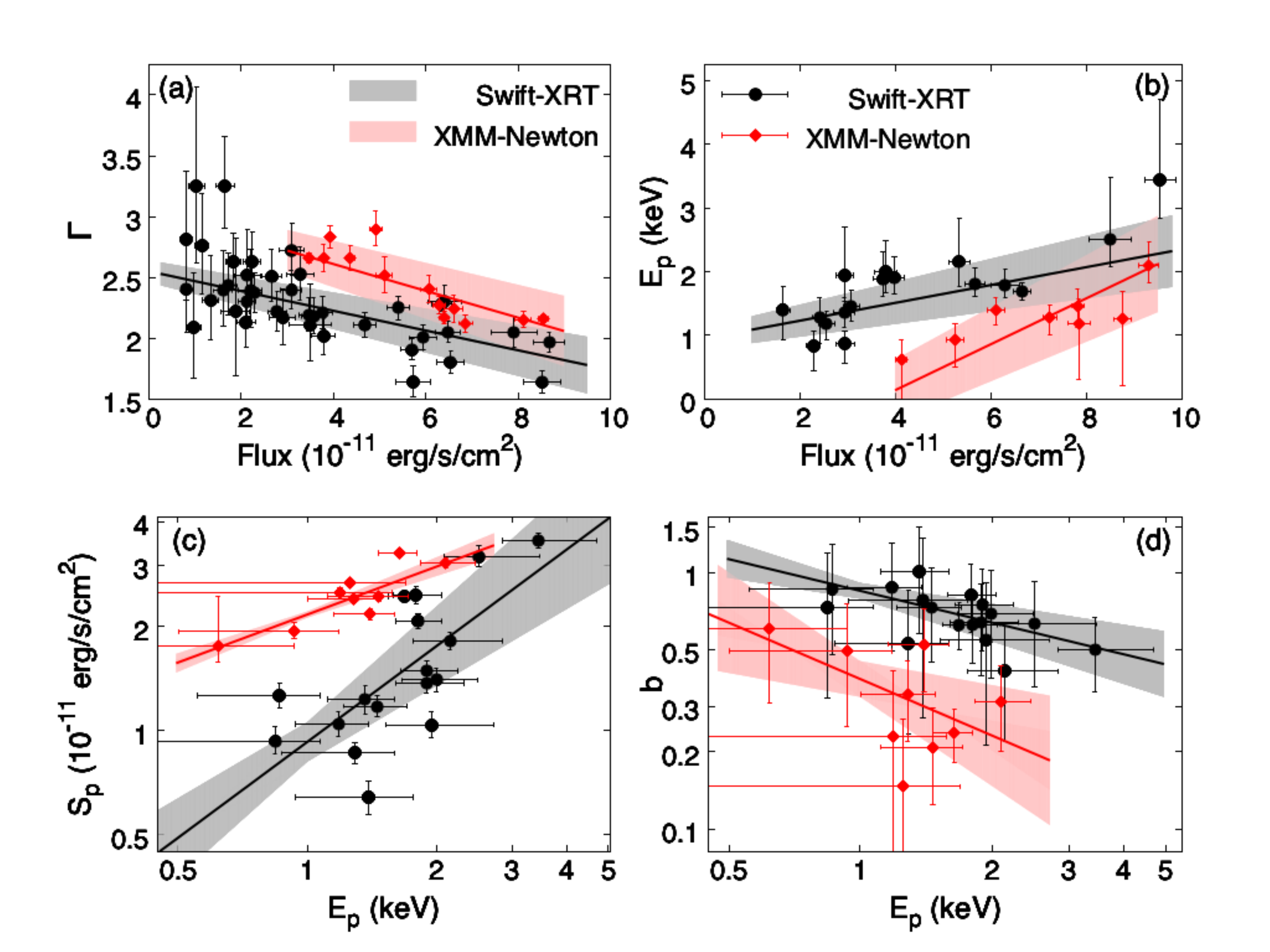}
\caption{Correlation between model parameters. {\it XMM-Newton} time-resolved data are indicated by red diamonds, 
and Swift data by black circles. (a) Power-law index versus count rate. (b) Peak energy ($E_{\rm p}$)
 versus count rate. (c) SED peak value ($S_{\rm p}$) versus SED peak energy ($E_{\rm p}$). 
(d) Spectral curvature $b$ versus peak energy ($E_{\rm p}$).}
\label{fig:xrate_index}
\end{figure*}

We perform correlation studies among the flux, PL and LP model parameters 
to gain insights on the radiation process (see Fig.~\ref{fig:xrate_index}).
For all cases linear correlation gives better fit statistics than 
constant correlation.
We calculate Pearson's Correlation Coefficient \citep{pearson}
along with standard deviation \citep{bowley28} for these model parameters 
using Python module {\it Scipy} \citep{scipy}. 
In Fig.~\ref{fig:xrate_index}(a), we compare the X-ray flux with the power-law index and find the following relation: 

\begin{align}
for\ & {\it XMM-Newton}: \nonumber \\
& \Gamma = (-0.11\pm0.02)\times({\rm flux})+(3.05\pm0.11)[\chi^2_\nu=3.09(11)], \\
for\ & {\it Swift-XRT}: \nonumber \\ 
& \Gamma =(-0.08\pm0.01)\times({\rm flux})+(2.55\pm0.09)[\chi^2_\nu=1.9(39)].
\end{align}

The correlation coefficient of the X-ray flux and power-law index are 
$r=-0.86 \pm 0.08$ (for {\it XMM-Newton} data) and $r=-0.55 \pm 0.11$ ({\it Swift} data) respectively. 
These results indicate a harder-when-brighter behavior. 

In Fig.~\ref{fig:xrate_index}(b), we compare the X-ray flux with the peak energy. 
The corresponding correlation coefficient is $r=0.81 \pm 0.13$ (for {\it XMM-Newton} data) 
and $r=0.83 \pm 0.08$ ({\it Swift data}), confirming the harder-when-brighter behavior. 
Indeed, we find for {\it XMM-Newton} and {\it Swift} the following relation: 

\begin{align}
for\ & {\it XMM-Newton}: \nonumber \\
& E_{\rm p} = (0.36 \pm 0.05)\times({\rm flux})-(1.30 \pm 0.28)[\chi^2_\nu=1.2(11)], \\
for\ & {\it Swift-XRT}: \nonumber \\ 
& E_{\rm p} = (0.14 \pm 0.04)\times({\rm flux})+(0.95 \pm 0.17)[\chi^2_\nu=1.0(16)].
\end{align}

Next, we study the relations between the peak energy ($E_{\rm p}$), SED peak value ($S_{\rm p}$) and spectral 
curvature $b$, in a similar manner as in \citet{2007A&A...466..521T}. For {\it XMM-Newton} and {\it Swift} data, 
we obtain

\begin{align}
for\ & {\it XMM-Newton}: \nonumber \\
& \ln{S_{\rm p}} = (0.46 \pm 0.06)*\ln{E_{\rm p}} + (0.77 \pm 0.02)[\chi^2_\nu=0.002(7)],\\
& \ln{b} = (-0.73 \pm 0.42)*\ln{E_{\rm p}} - (0.95 \pm 0.16)[\chi^2_\nu=0.01(7)],\\
for\ & {\it Swift-XRT}: \nonumber \\ 
& \ln{S_{\rm p}} = (0.92 \pm 0.19)*\ln{E_{\rm p}} - (0.07 \pm 0.13)[\chi^2_\nu=0.003(16)],\\
& \ln{b} = (-0.41 \pm 0.14)*\ln{E_{\rm p}} - (0.16 \pm 0.07)[\chi^2_\nu=0.03(16)].
\end{align}

The unit of $S_{\rm p}$ is 10$^{-11}$~erg~cm$^{-2}$~s$^{-1}$ and $E_{\rm p}$ is in keV.

The Pearson's correlation coefficient between $\ln{S_{\rm p}}$ and $\ln{E_{\rm p}}$ is $r=0.86 \pm 0.10$ 
(for {\it XMM-Newton} data) and $r=0.71 \pm 0.12$ ({\it Swift} data), showing strong positive correlation. Within the context 
of the synchrotron emission from one dominant component, $S_{\rm p}$ depends on $E_{\rm p}$ as: 
$S_{\rm p}\propto E_{\rm p}^\alpha$. The value of $\alpha\sim$0.5--0.9, we find here is smaller than unity, 
indicating that the spectral change should be caused by variation of the electron average energy ($\alpha=1.5$),
 or to the magnetic field change ($\alpha=2$), but not due to change in the beaming factor 
\citep[$\alpha=4$;][]{2007A&A...466..521T}. The result is shown in Fig.~\ref{fig:xrate_index}(c).
The correlation between $\ln{b}$ and $\ln{E_{\rm p}}$ is significant ($r=-0.50 \pm 0.29$) for {\it XMM-Newton} data but not for 
{\it Swift} data ($r=-0.66 \pm 0.14$) 
\citep[with $b\propto E_{\rm p}^{-0.45}$, and such a negative correlation is expected in 
statistical or stochastic acceleration;][one should note that {\it Swift} data were taken 
over a long time span (i.e., more than a year) while the {\it XMM-Newton} observation was 
taken within a day]{2007A&A...466..521T}. 
Therefore, it is plausible that different physics is driving the spectral shape (referring here to the peak 
energy and curvature) at different time scales. The result is shown in Fig.~\ref{fig:xrate_index}(d).

In summary, owing to the high sensitivity of {\it XMM-Newton}, we have found the rapid X-ray variation from 
\fermi\ with timescale down to $\sim$1~hour, and a hardening X-ray spectrum following the rise of the X-ray flux. 
Both of these findings support a blazar scenario, in which the X-ray emission is dominated by synchrotron emission 
of a relativistic jet component.

\subsubsection{Other X-ray observations}
{\it MAXI-GSC} 2--20 keV light curve with 1 day time bin was obtained from a circular region at the best source position 
with 1.6$\arcdeg$ radius from {\it MAXI} online data reduction system\footnote{{\it MAXI} on-demand queries, 
\url{http://134.160.243.88/mxondem/}}. Some excess can be seen in the light curve during and shortly after the two 
major flares in May and August 2017, respectively.

{\it ROSAT-PSPC} observed this position of sky for a total exposure $\sim$480~s during the {\it all-sky survey (RASS)}. Since 
there is no detection in this position, and taking 5 counts as a minimum for a detection, the upper limit of the count 
rate is approximately 0.01. The upper limit of energy flux is taken to be about $10^{-13}$~erg~cm$^{-2}$~s$^{-1}$ 
\citep[2RXS;][]{2rxs}. {\it Swift-XRT} observations thus revealed a peak X-ray flux more than $10^3$ higher than this upper 
limit. These values are indicated by dashed lines in the {\it XRT} count rate and light curve panels in Fig.~\ref{fig:mwl_lc}.

\section{UV, Optical, and IR properties}
\subsection{UV and optical photometric measurements}
\label{sec:uvot_phot}

\begin{figure*}
\includegraphics[scale=0.27,angle=270]{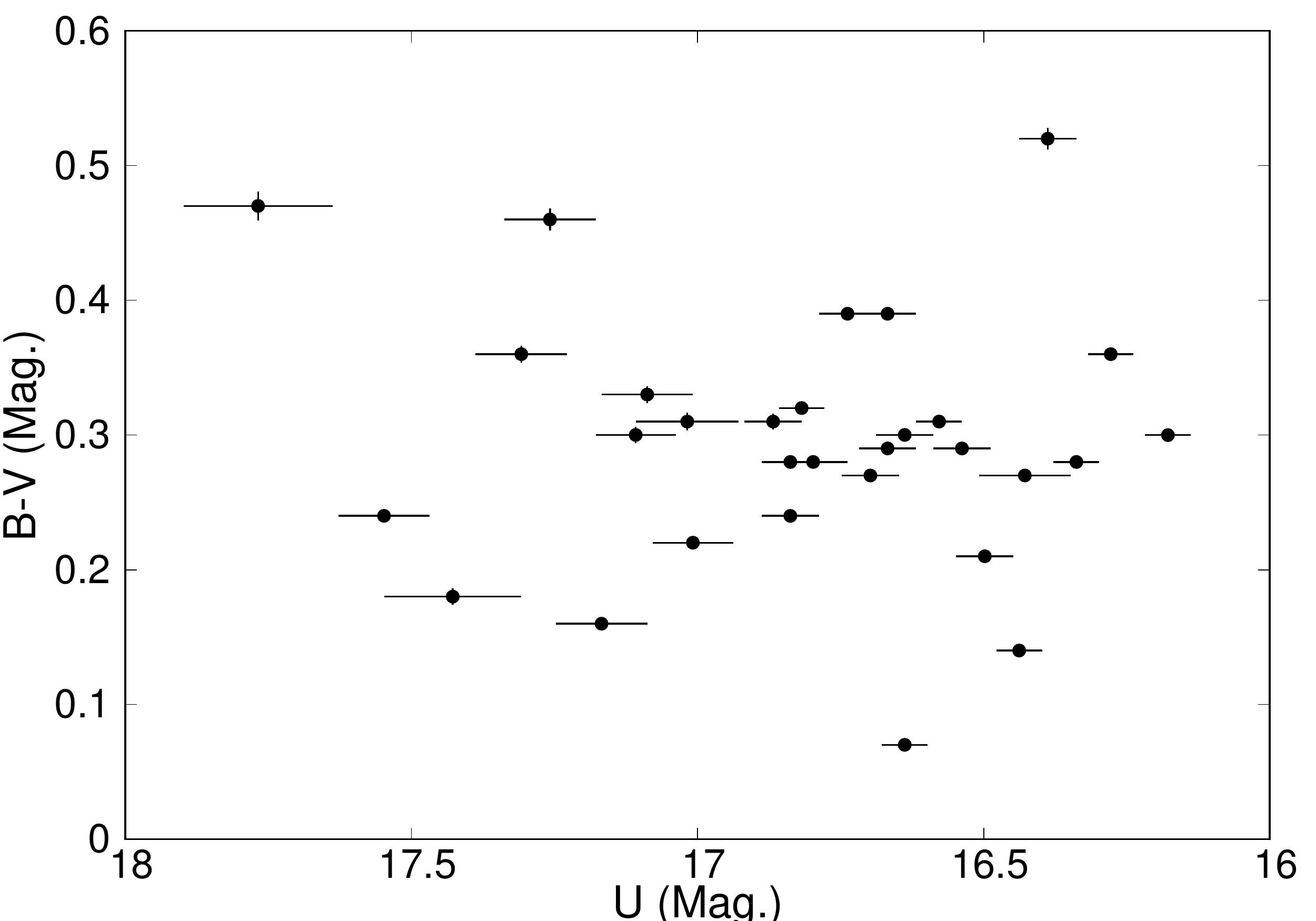}
\includegraphics[scale=0.27,angle=270]{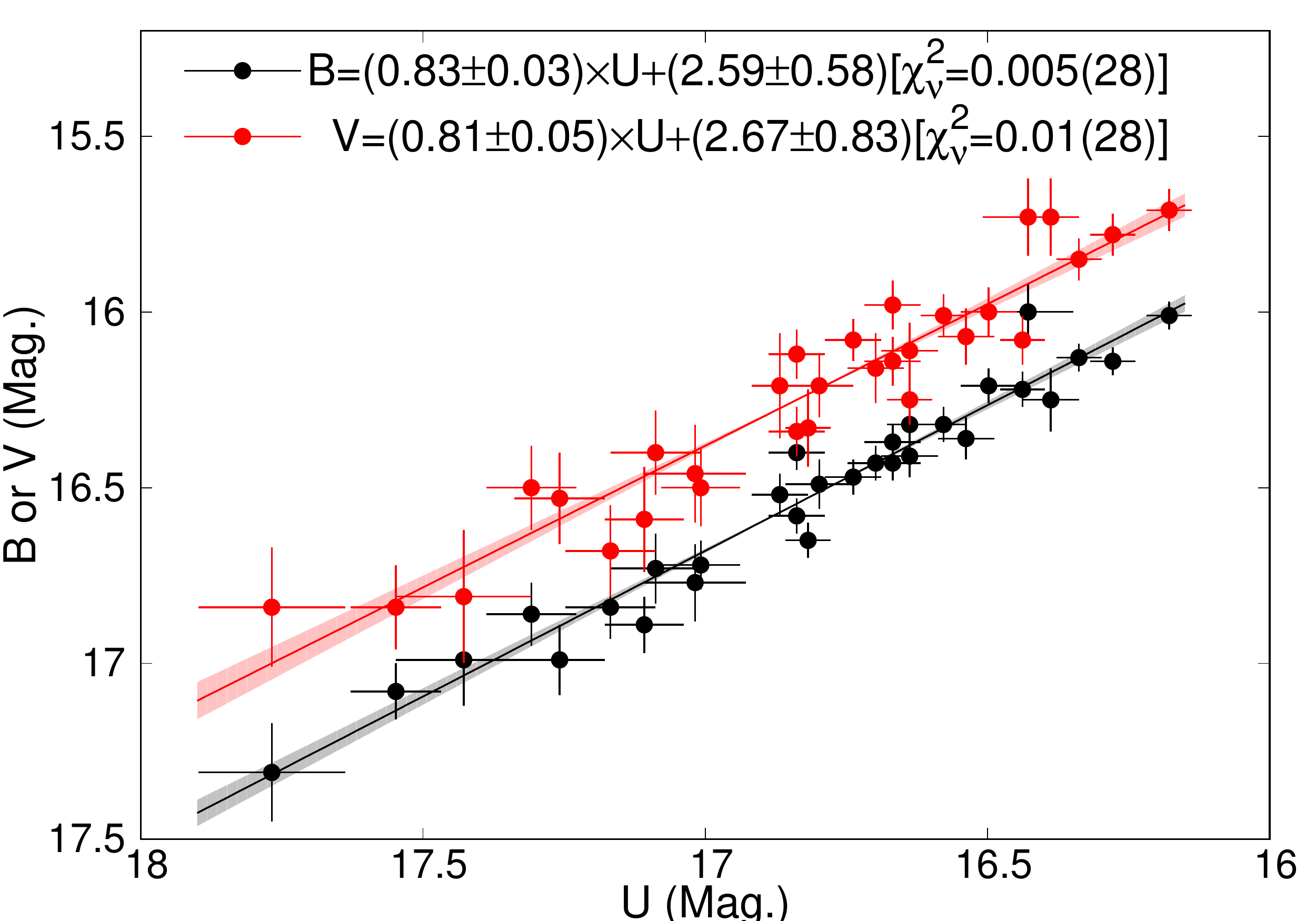}
\caption{Left: the flux-color plot for optical observations. The x- and y- axis is the U-band magnitude and 
\bv color index, respectively. Right: B- or V-band magnitude versus U-band magnitude for various optical 
flux. The black straight line has a slope of $0.83 \pm 0.03$, and the red line a slope of $0.81 \pm 0.05$. 
Galactic extinction is corrected for in plotting these figures (see Section.~\ref{sec:uvot_phot}).}
\label{fig:ubv_correlate}
\end{figure*}

For {\it Swift-UVOT} data reduction, all extensions of sky images were stacked with {\it uvotimsum}.
The source magnitudes were derived with 3-$\sigma$ significance level      
from the circular region of 5$\arcsec$ radius in the best source position
of the stacked sky images from all the filters with {\it uvotsource}. The background was estimated from an 
annular region in the same position with radii from 10$\arcsec$ to 20$\arcsec$.
The {\it Swift-UVOT} light curves (extinction not corrected) of different filters are plotted in Fig.~\ref{fig:mwl_lc}. 
To check any color change, we also applied interstellar de-reddening on the U, B, and V-magnitudes. The value of 
extinction was estimated using the web-based calculator maintained by the NASA/IPAC Infrared Science 
Archive\footnote{\url{http://irsa.ipac.caltech.edu/applications/DUST/}} \citep{S_F11}, and the 
corrected values are shown in Fig.~\ref{fig:ubv_correlate}. It can be seen that there is no color change against 
different U-band flux. In particular, no bluer-when-brighter behavior is seen.

On 2018 February 21, {\it XMM-OM} observed \fermi\ in FAST mode for 12 exposures with different filters.
For {\it XMM-OM} data reduction, all exposures of sky images are extracted with {\it omfchain}.
The source magnitudes are derived with 3-$\sigma$ significance level      
from the circular region of 5$\arcsec$ radius in the best source position
of the stacked sky images from all the filters with {\it omdetect}. 
The background is estimated from an annular region in the same position with radii from 10$\arcsec$ to 20$\arcsec$.
The absolute magnitudes obtained from the analysis are shown in Table.~\ref{xtab3}.

\begin{table*}
\begin{center}
\caption{{\it XMM-Newton OM} analysis results of \fermi~on 2018 February 21 \label{xtab3}}
\begin{tabular}{cccc}
\hline
Exposure & Exposure & OM  & Absolute \\
identifier & (s) & Filter & magnitude \\
\hline
S014 & 4400 & V & 16.96$\pm$0.02 \\
S015 & 4400 & U & 16.86$\pm$0.01 \\
S016 & 4400 & B & 17.69$\pm$0.01 \\
S017 & 4400 & B & 17.63$\pm$0.01 \\
S018 & 4400 & UVW1 & 16.57$\pm$0.02 \\
S019 & 4400 & UVW1 & 16.48$\pm$0.01 \\
S020 & 4400 & UVM2 & 16.71$\pm$0.04 \\
S021 & 4400 & UVM2 & 16.80$\pm$0.04 \\
S022 & 4400 & UVM2 & 16.71$\pm$0.04 \\
S023 & 4400 & UVW2 & 16.87$\pm$0.07 \\
S024 & 4400 & UVW2 & 16.75$\pm$0.06 \\
S025 & 3340 & UVW2 & 16.89$\pm$0.08 \\
\hline
\end{tabular}
\end{center}
\end{table*}

In summary, as seen in Fig.~\ref{fig:mwl_lc}, the evolution of the optical emission from \fermi\ is independent 
of that of the high-energy (i.e., X/\gr) emission. We did several tests \citep[including a zDCF code][]{zdcf}
and did not find a correlation between \gr\ or X-ray flux with optical/NIR flux. In particular, the average optical 
flux in 2018 is higher than that in 2017, but the average X/\gr\ flux is higher in 2017 than in 2018. This may be 
due to two emission regions not directly related to each other.

\subsection{Long-term Optical and Mid-infrared light curves}
\label{sec:oir}

A comprehensive examination of the long-term variability of \fermi\ in every other available band is 
helpful for us to understand its nature. We checked its optical and mid-infrared (MIR) light curves 
as shown in Fig.~\ref{fig:oirlc}. The optical ($V$-band) data are retrieved from public searching 
server of {\it Cataline Real-Time Transient Survey}\footnote{\url{http://nunuku.caltech.edu/cgi-bin/getcssconedb\_release\_img.cgi}} 
\citep[{\it CRTS};][]{Drake09} and {\it All-Sky Automated Survey for Supernova} 
\citep[{\it ASAS-SN}][]{Shappe14, Kochanek17}\footnote{\url{https://asas-sn.osu.edu/}}.
Although with large photometric errors, a long-term variation is clearly visible,
indicative of an AGN. After measurement errors are taken into account, the CRTS variability
amplitude ($\Delta V$) is $\sim0.07$ mag \citep[e.g., Equation~6 in][]{sesar07}. 
The {\it ASAS-SN} data possess even larger errors due to its shallow survey depth.
Despite that, we can still see a significant (i.e., at the 4-$\sigma$ level) brightening in the latest two epochs ($\sim0.7$ mag).

In addition to the ground-based optical time-domain surveys, 
the {\it Wide-field Infrared Survey Explorer} \citep[\it{WISE}][]{Wright10, Mainzer14} 
has scanned a specific sky area every half year at 3.4 and 4.6~$\mu$m (labeled {\it W1} and {\it W2}) since 2010 Feb 
(except for a gap between 2011 Feb and 2013 Dec) and thus yielded 12-13 times of observations for each object up to now.
We downloaded all of the public {\it WISE} data of \fermi~up to the end of 2018 July,
distributing over 12 epochs at intervals of half year.
For each epoch, there are typically 12 individual exposures within one day.
Hence the {\it WISE} database allows us to study both its long-term and intra-day MIR variability.
First, we binned the data every half year \citep[as we have done in][]{Jiang16, Dou2016}, 
which displays an obvious and continuous trend of brightening since 2017 February. 
The latest exposures taken in 2018 July has brightened by $\sim$1.3 and $\sim$1.5 magnitudes in 
{\it W1} and {\it W2}, respectively, in comparison with two years earlier; such an increase is even larger than in the optical band.
We have also tried to explore the possibility of intra-day variability in each epoch
following \citet{Jiang12, Jiang18}, which may provide a direct evidence for the jet toward us.
Nevertheless, the short-timescale variability is insignificant.

In summary, there are long-term variations in both optical and MIR bands, that are indicators of past AGN activity. 
Moreover, both bands show a trend of recent brightening, especially in 2018 when the high-energy emission goes down to a lower state.

\subsection{Optical spectroscopy}
To look for any spectral feature in optical, we obtained three spectra in 2017 and 2018. We carried out an observation 
using the {\it IMACS (f/2) spectrograph} on the {\it 6.5-m Magellan telescope} on 2017 September 7, with a total exposure time of 800s. 
Two standard stars and He--Ne--Ar lamp spectra were taken before and after the exposure for flux and wavelength calibration. 
The raw two-dimensional data reduction and spectral extraction were accomplished using standard routines in {\it IRAF}. 
To extract the nuclear spectra, we used the {\it APALL} task and chose an aperture of 2$\arcsec$.

We also performed a spectroscopic observation of \fermi~by the {\it Yunnan Faint Object Spectrograph and 
Camera (YFOSC)} on the {\it 2.4m telescope}, located at the {\it Lijiang Station of Yunnan Observatories 
(longitude = 100$\arcdeg$01$\arcmin$51$\arcsec$, latitude = 26$\arcdeg$42$\arcmin$32$\arcsec$N, altitude = 3193 m)} 
of the Chinese Academy of Sciences on 2018 February 28. Grism \#14 of YFOSC, which has a resolution of 1.67\AA~pixel$^{-1}$ 
and wavelength coverage of 3200--7500\,\AA, was used. Given the seeing conditions, we employed a slit with a width of $2\farcs5$. 
The total exposure time is 3300~s to achieve a high signal-to-noise (S/N) ratio. The spectroscopic data were reduced following 
the standard procedures using {\it IRAF}, including bias and flat correction, cosmic ray rejection, spectrum extraction, wavelength 
calibration, and flux calibration. When extracting the spectrum, the aperture was selected to reach 2\% of the peak value to 
include most of the light from the source; a good S/N ratio can therefore be obtained. The emission line of He-Ne lamp was 
used for wavelength calibration. The aperture of the lamp spectrum was identical to the aperture of the source, ensuring that 
function between wavelength and the position corresponds to the aperture of the source. {\it BD+33d2642} is used as the spectroscopic 
standard star to calibrate the flux of the object. Considering the airmass of the standard star and the extinction coefficient 
at Lijiang Station, the sensitivity function can be determined by using the counts and the flux at each wavelength for the standard star. 
Then the sensitivity function is applied to \fermi\ to convert the counts back to flux for \fermi. The airmass of the object, 
which is different from the standard star, is also considered.

On 2018 May 11, we obtained a medium resolution (R$\sim$2000) spectrum using the {\it Kast double spectrograph} 
(consisting of red and blue channels) on the {\it 3-m Shane telescope} at the {\it Lick Observatory}. We used the
600/4310 grism on the blue size and 600/5000 grating on the red side with a wavelength coverage approximately 
3300--5500\AA~and 5500--8000\AA. We apply a $1\farcs5$ slit aligned at parallactic angle for the observation 
with a 30-minute exposure for both channels. The flux calibration is based on the spectrophotometric standard star {\it Feige~67}.

\begin{figure*}
\centering
\includegraphics[width=0.6\paperwidth]{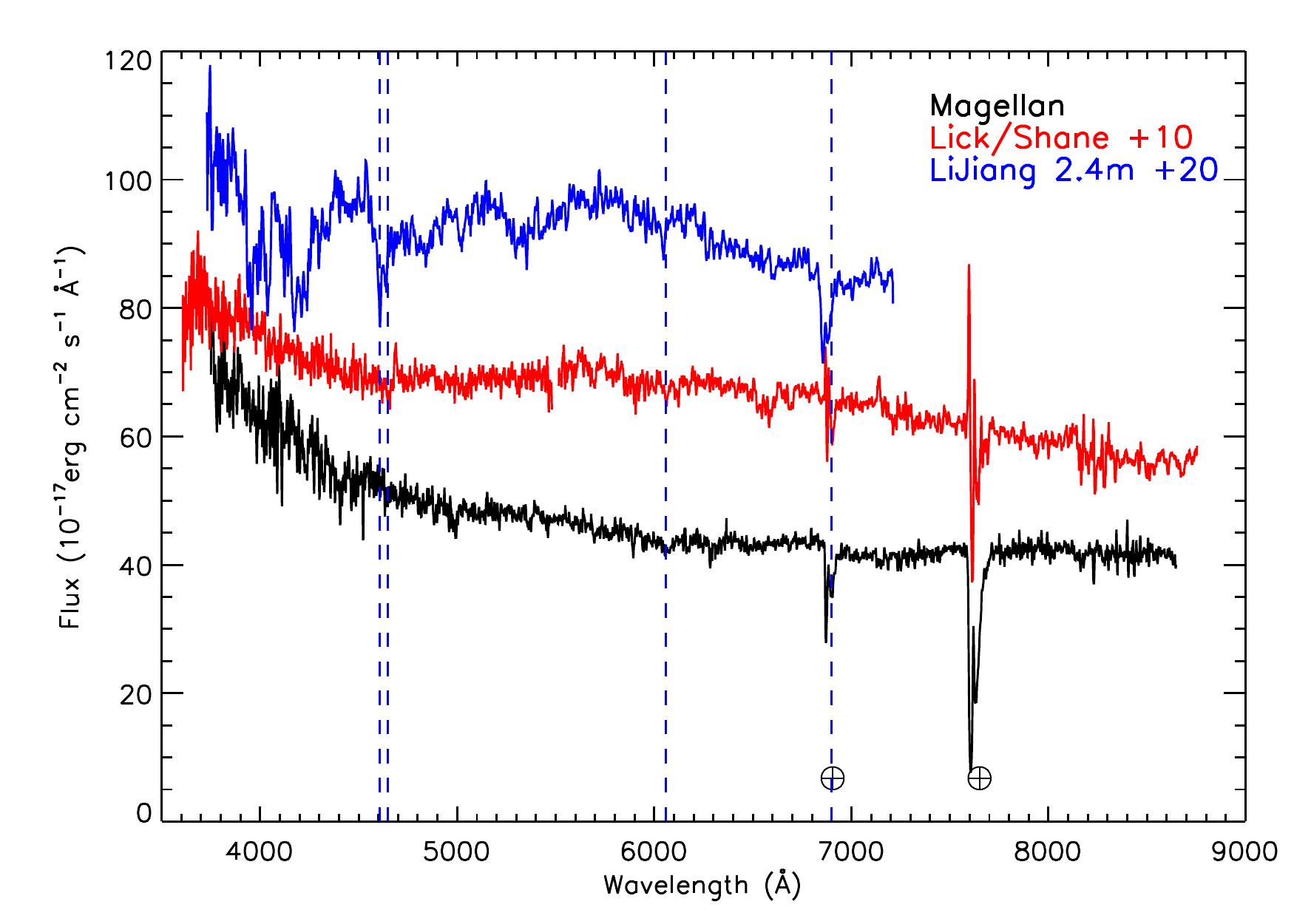}
\caption{The Lijiang 2.4m spectrum taken on 2018 February 28 (where two possible broad absorption lines (BALs) at around 
4000\AA~and 4200\AA~are evident), is compared with those taken by the {\it Magellan 6.5m telescope} on 2017 September 7 
and the {\it Lick-Shane telescope} on 2018 May 12. The two crosses mark the position of telluric absorption. Indicated by 
vertical dashed lines are those of CaII and K, MgI and NaI at z=0.171 (from left to right), but the NaI position 
is very close to a telluric feature and thus the identification is only tentative. The Magellan spectrum is smoothed 
by 3 pixels and the {\it Lijiang} and {\it Lick-Shane} spectrum by 5 pixels for display purpose.}
\label{fig:opt_spec}
\end{figure*}

The {\it Magellan} and {\it Lick} spectra obtained are largely featureless with only weak absorption lines (Fig.~\ref{fig:opt_spec}). 
These spectra are consistent with \fermi\ being a BL Lac object. Along with {\it Lijiang} observation, these three optical 
spectra (taken at times shown by the dashed lines of Fig.~\ref{fig:mwl_lc}) has shown strong variation which is not correlated 
with the X\&\gr\ variation.  
We also checked that the stellar absorption lines from host galaxy \citep[e.g. CaII, K, MgI and NaI][]{paiano17} 
in the {\it Lijiang} spectrum  have a redshift consistent with 0.17 \citep{Chornock17,Bruni18}.
The {\it Lijiang} spectrum was taken on 2018 February 28. 

Surprisingly, in the blue end, there are two unidentified BAL-like features at around 4000~\AA~and 4200~\AA. 
If it is true, such BAL feature would indicate fast gas outflows blocking the line of sight towards the unknown 
optical source. This phenomenon is often observed in quasars \citep{weymann91}, and has been seen in at least 
one BL Lac object, PKS B0138-097 \citep{zhang11}. 
Such BAL-like feature does appear just a week after the {\it XMM-OM} observation (21 February 2018, MJD 58170) 
when the optical spectrum (as seen by the magnitudes in different filters) is very blue, as compared to those taken in other epochs 
(Fig.~\ref{fig:mwl_lc} fifth panel); thus the BAL-like feature is seen during a unique optical color-changing state. 
The Lijiang observation is performed at the second part of the night before the early morning, sometime, 
the observation condition changes quickly due to the frog or wet air. However, according to the note by the Lijiang observer, 
there is no evidence of either quick air change, or instrumental malfunction during the observation time. \fermi\ was observed at 
2018-02-28 20:41:55.601 for 3300s, and the airmass is 1.359291. The standard star bd332642 was observed at 2018-02-28 20:33:05.119 
for 200s, and the AIRMASS is 1.12949. Of course, we could not exclude any undiscovered problems related to the instrument.  

\section{{\it RATAN 600-meter} radio observations}

The measurements of the fluxes were obtained with the {\it RATAN-600m} radio telescope in transit mode by observing simultaneously 
at 1.2, 2.3, 4.8, 8.2, 11.2, and 21.7 GHz. The observations were carried out during October and December 2017, and January, 
February-April and July 2018. The parameters of the antenna and receivers are listed in Table.~\ref{tab:radiometers}, 
where $f$ is the central frequency, $\Delta f$ is the bandwidth, $\Delta F$ is the flux density detection limit per beam,
 and $BW$ -- beam width (full width at half-maximum in RA). The detection limit for the {\it RATAN} single sector is approximately 
5~mJy (the time of integration is 3~s) under good conditions at the frequency of 4.8~GHz and at an average antenna elevation. 
We averaged the data of observations for 2-25 days in order to get a reliable values of the flux density. 
Data were reduced using the {\it RATAN} standard software {\it FADPS (Flexible Astronomical Data Processing System)} reduction package 
\citep{1997ASPC..125...46V}. The flux density measurement procedure is described by 
\citet{2012A&A...544A..25M, 2014arXiv1410.2835M, 2016AstBu..71..496U, 2017AN....338..700M}.
The following flux density calibrators were applied to obtain the calibration coefficients in the scale by
\citet{1977A&A....61...99B}: 3C48, 3C147, 3C161, 3C286 and NGC7027. We also used the traditional {\it RATAN}
flux density calibrators: J0237$-$23, 3C138, J1154$-$35, and J1347$+$12. Measurements of some calibrators were corrected
for linear polarization and angular size, following the data from \citet{1994A&A...284..331O}
and \citet{1980A&AS...39..379T}. The systematic uncertainty of the absolute flux scale (3-10{\%} at different {\it RATAN} frequencies) 
is not included in the flux error. The total error in the flux density includes the uncertainty of {\it RATAN} calibration curve
and the error in the antenna temperature measurement.

\begin{table*}
\centering
\caption{\label{tab:radiometers}Parameters of the {\it RATAN-600m} antenna and radiometers}
\centering
\begin{tabular}{rlcr}
\hline
\hline
 $f$ & $\Delta$$f$ & $\Delta$$F$ &  FWHM$_{\rm x}$ \\
  GHz    &   GHz       &  mJy beam$^{-1}$ &  arcsec \\
\hline
 $21.7$ & $2.5$  &  $50$ & 11 \\
 $11.2$ & $1.4$  &  $15$ & 16 \\
 $8.2$  & $1.0$  &  $10$ & 22 \\
 $4.8$  & $0.6$  &  $5$  & 35 \\
 $2.25$ & $0.08$ &  $40$ & 80 \\
 $1.28$ & $0.08$ &  $200$& 110 \\
\hline
\end{tabular}
\end{table*}

\begin{figure*}
\centering
\includegraphics[width=0.6\paperwidth]{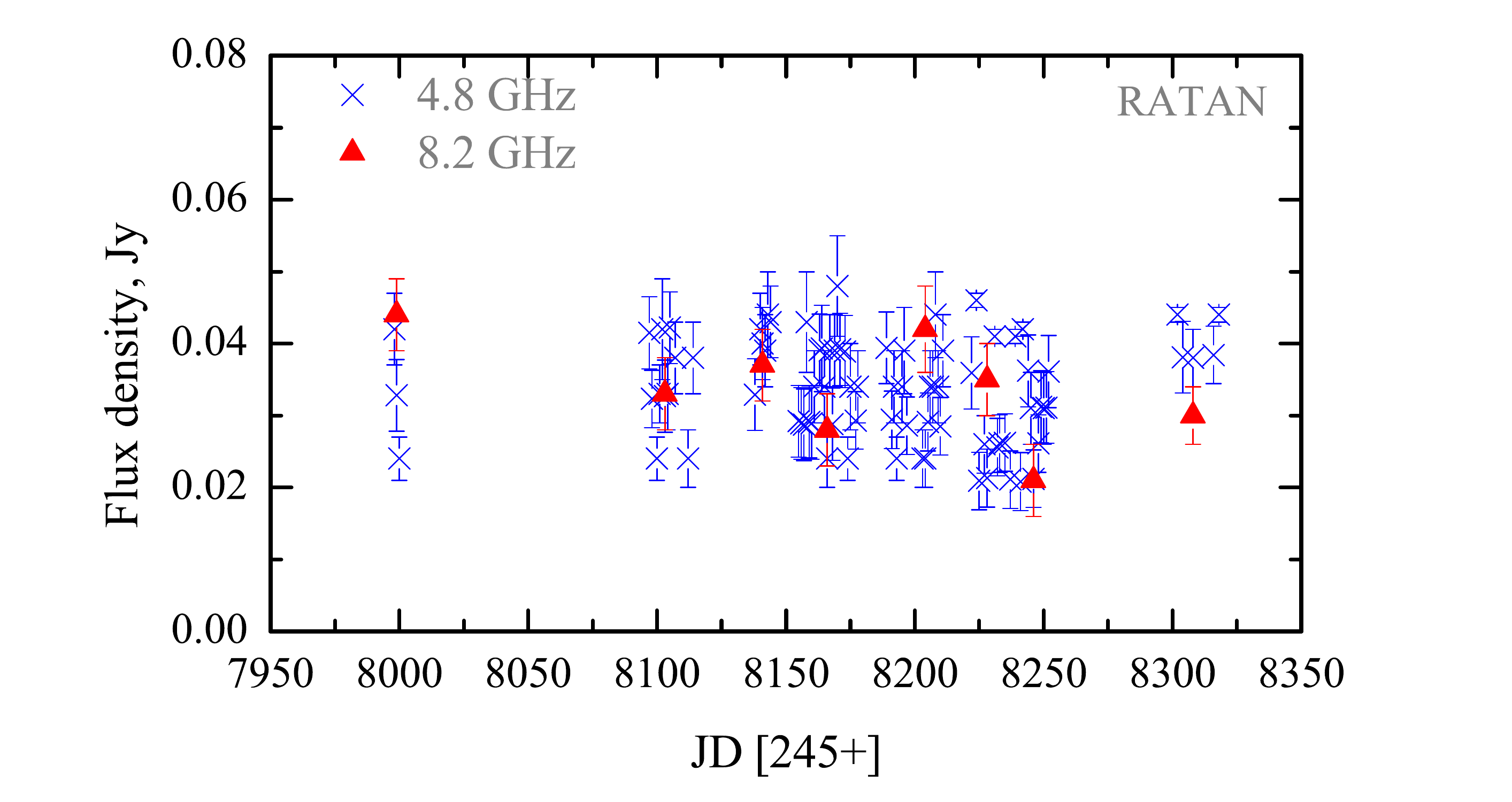}
\caption{The light curves for \fermi\ at 4.8 and 8.2 GHz, obtained by {\it RATAN-600m} observations.}
\label{fig:ratan_lc}
\end{figure*}

The radio emission was detected at 4.8~GHz and 8.2~GHz only. We measured the flux density at 4.8 GHz for each single scan. 
At the frequency of 8.2 GHz we used all scans in each observation epoch (Fig.~\ref{fig:ratan_lc}) to get average flux density.
We did not find any significant variation of the flux density in the 2017--2018 measurements.
The average flux densities at 4.8 and 8.2 GHz and number of observations in each month are presented in Table.~\ref{tab:ratan_flux}.

\begin{table*}
\caption{\label{tab:ratan_flux}The monthly-average flux densities of \fermi~obtained with the {\it RATAN-600m}.}
\centering
\begin{tabular}{lrrr}
\hline\hline
Epoch & $N_{\rm obs}$ & $S_{\rm 8.2GHz}$ & $S_{\rm 4.8GHz}$ \\
      & &   (Jy)      &  (Jy)           \\
\hline
September 2017 & &  & \\
2457998-2458000 & 3  & 0.044$\pm$0.005  &  0.039$\pm$0.002   \\
\hline
December 2017  & &   & \\
2458097-2458114 & 11 & 0.033$\pm$0.005  &  0.038$\pm$0.003   \\
\hline
January 2018  & &  & \\
2458138-2458144 & 6 & 0.037$\pm$0.005  &  0.040$\pm$0.002   \\
\hline
February 2018  & &  & \\
2458155-2458178 & 21 & 0.028$\pm$0.005  &  0.039$\pm$0.002   \\
\hline
March    2018   & &  & \\
2458189-2458211 & 16 & 0.042$\pm$0.006  &  0.032$\pm$0.002   \\
\hline
April    2018   & &  & \\
2458222-2458239 & 11 & 0.035$\pm$0.005  &  0.030$\pm$0.002   \\
\hline
May    2018   & &  & \\
2458240-2458252 & 10 & 0.021$\pm$0.005  &  0.031$\pm$0.002   \\
\hline
July     2018   & &  & \\
2458302-2458318 & 5  & 0.030$\pm$0.004  &  0.040$\pm$0.003   \\
\hline
\end{tabular}
\end{table*}

\section{Discussion}
\label{sec:discuss}
\subsection{Is \fermi\ a blazar?}

The most enigmatic behavior of \fermi\ is its recent `turn-on' high-energy emission (X-rays and \grs) for about a year, 
while it remained quiescent for the past decade or so (see Fig.~\ref{fig:oirlc}). During this `turn-on' state, 
the \gr\ flux shows variabilities with a minimum time scale down to weeks. \citet{Bruni18} suggest that \fermi\ is a 
BL Lac object. Blazars are well known sources that show variability at all time scales from decades down to intra-day (i.e., IDV). 
And indeed, with {\it XMM-Newton}, who has a much better sensitivity than {\it Fermi}, we have discovered $<1$~hour 
variation during a GeV low state. 

If \fermi\ is a previously unknown blazar, its high-energy flux is constrained to be very low by all-sky monitors 
like {\it Fermi-LAT} and {\it MAXI} for around a decade before 2017, as well as {\it ROSAT} in the 1990s. 
The high-energy flares that began in May 2017 thus indicates a high-state never seen before for \fermi. 
Particularly in \grs, \fermi\ remains in the quiescent 
state for nearly 9 years, which is a rather long period for a {\it Fermi} blazar. When comparing the \gr\ spectrum of \fermi\ with 
sources in the second {\it LAT} AGN catalog \citep[2LAC;][]{second_lac}, the mean photon index of the three major categories of 
\gr\ BL Lac objects, high-synchrotron-peak (HSP), intermediate-synchrotron-peak, and low-synchrotron-peak, is 1.84, 2.08, and 2.32, 
respectively. With an average photon index of about 1.7, \fermi\ (during its flares) has a spectral index at \grs\ well within that 
of HSP BL Lac objects.

In both the low and flaring state, the SED of \fermi\ remains a typical blazar SED, which consists mainly of a 
synchrotron peak in X-rays and an inverse-Compton peak in \grs. During the high state in May 2017, 
the \gr\ peak indicates \fermi\ to be a high-frequency-peaked BL Lac object.
The changing X-ray photon index of \fermi\ is around $\Gamma\sim2.0$ during the major flaring period. 
i.e., $1.63\pm0.13$ on MJD 57936, $1.67\pm0.0.07$ on MJD 57991, and $1.63\pm0.09$ on MJD 58009, 
making \fermi\ to be an extreme blazar candidate (at times) based on the synchrotron peak frequency \citep{Abdo10,Fan16}.
Extreme high-frequency-peaked BL Lac objects (EHBLs), like 1ES~0229+200 and 1ES~1101-232, are blazars with 
a very high synchrotron peak \citep[c.f., $>$1~keV;][]{Costamante01} 
and usually exhibit exceptionally hard TeV spectra, and they are good probes of the Extra-galactic Background Light (EBL) 
and Extra-galactic Magnetic Field (EGMF). Yet the sample of extreme blazars remains small \citep{Costamante18}. 
Mkn~501 behaved like an EHBL throughout the 2012 observing season, with low and high-energy components peaked above 
5 keV and 0.5 TeV, respectively. This suggests that being an EHBL may not be a permanent characteristic of a blazar, 
but rather a state which may change over time \citep{Pian98, ahnen18}. 
Future X-ray/TeV measurements may help us, not just to further probe the synchrotron/IC peak of \fermi, but also 
to unveil its rapid variations, even during a GeV low state.

\subsection{A simple SSC model for this blazar candidate -- \fermi}
\label{model}
\begin{figure*}
\centering
\includegraphics[width=6.5cm]{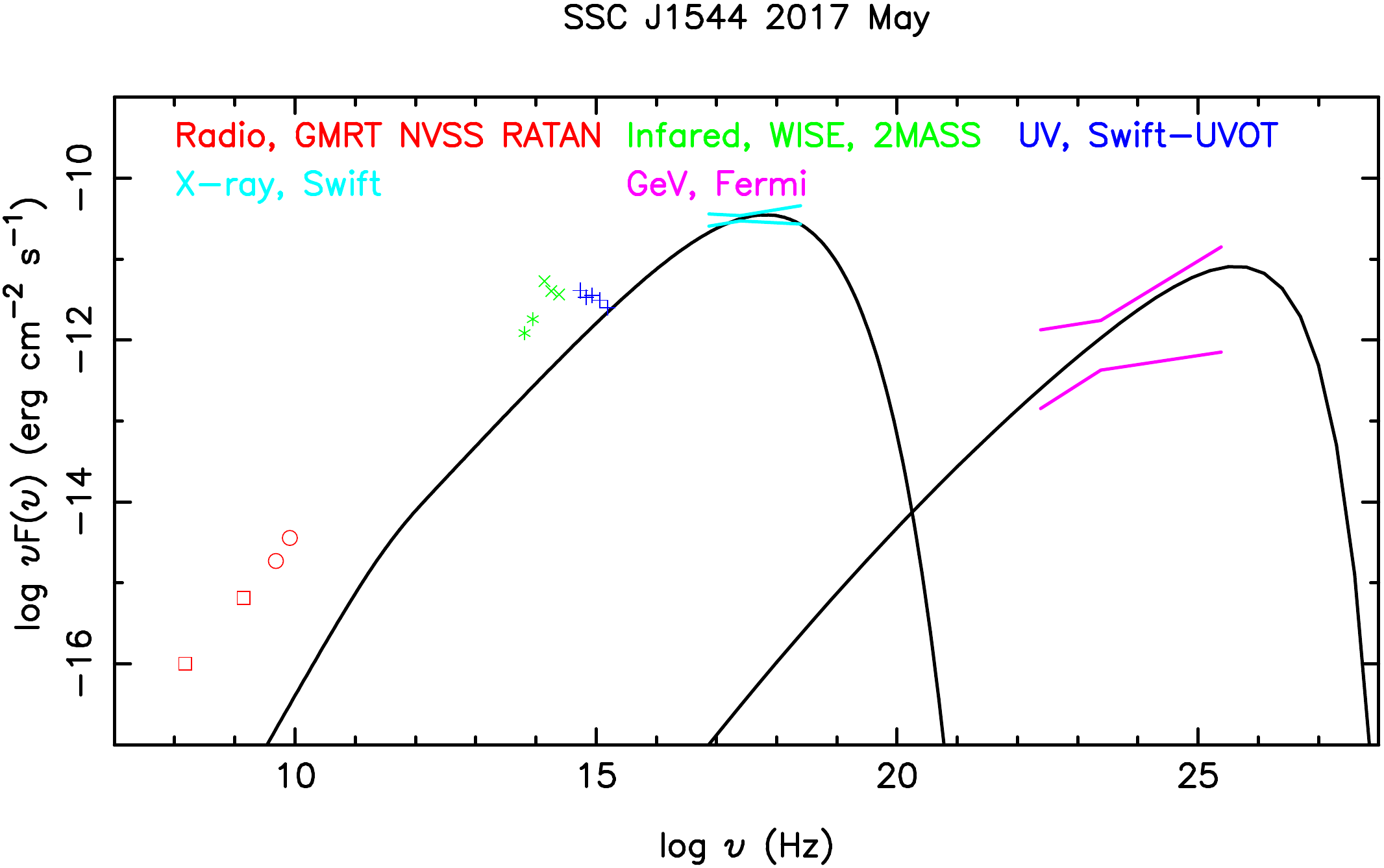}
\includegraphics[width=6.5cm]{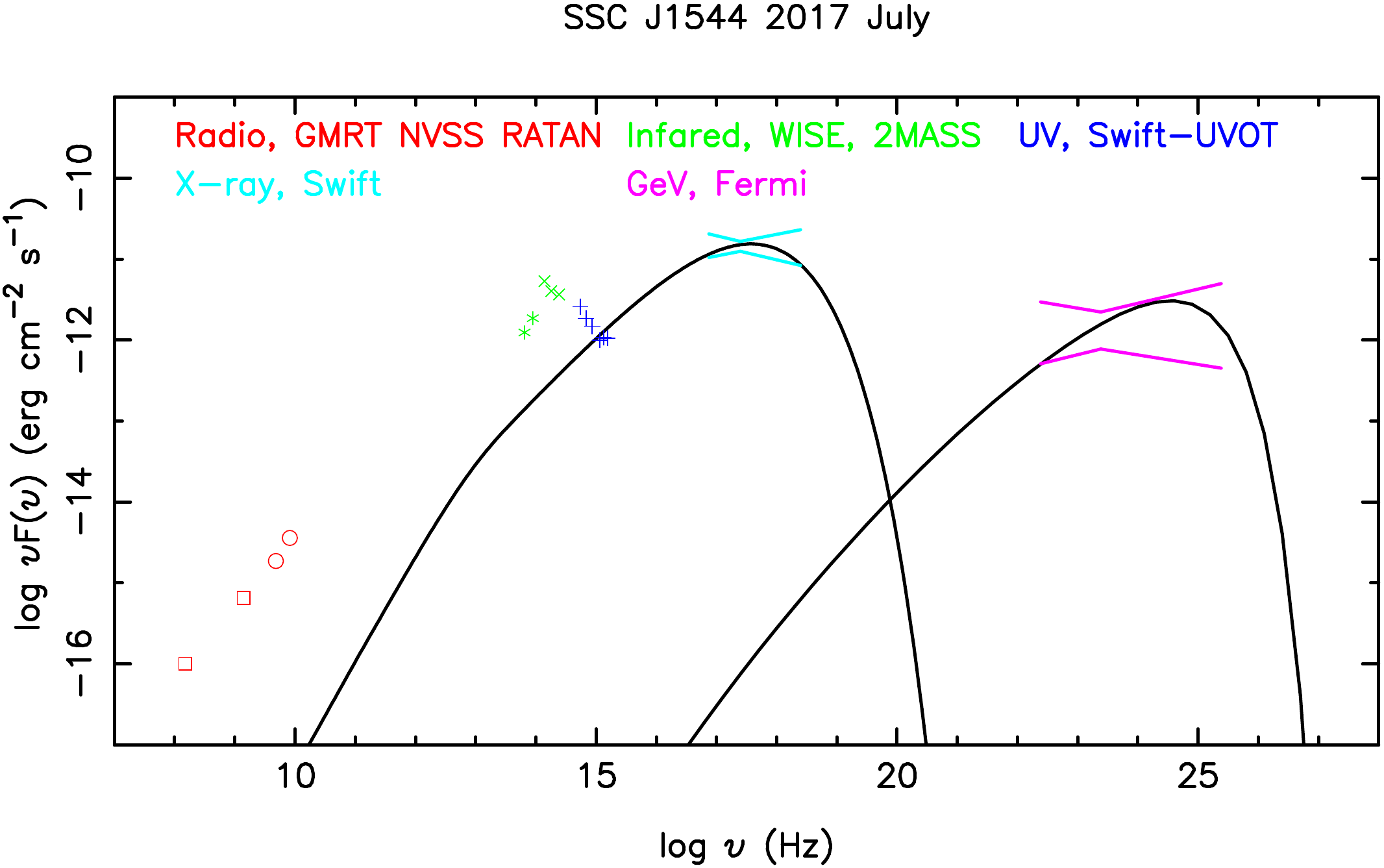}
\caption{The SED for two representative dates: 2017 May 26 (left panel for the first flare) and 2017 July 1 
(right panel, between first and second flare). The model lines are from a one-zone SSC model described in the text. 
The {\it Fermi} spectra were derived from 2017 May 25--27 and 2017 June 30--July~2, respectively. The 4.8 and 8.2 GHz data 
are from {\it RATAN-600m} observations taken in September 2017. 
Galactic extinction is not corrected for the UV/optical/NIR data shown.}
\label{sed}
\end{figure*}

Following the hypothesis that \fermi\ is a blazar, we employ a simple Synchrotron Self-Compton (SSC) model 
to estimate the physics in \fermi. This SSC zone could be either from the main jet in a typical blazar scenario, 
or from a mini-jet in a misaligned blazar scenario. Since the source is varying by a large factor, 
spectral energy distribution (SED) are shown for two representative dates: 2017 May 26 (during the first flare) 
and 2017 July 1 (between the first and second flare), as shown in Fig.~\ref{sed}. It has a broad-band SED 
consistent with a BL Lac object.

During the SSC fitting of the SED of \fermi, the first constraint comes from the relatively low UV emission 
when compared to the X-ray emission, clearly the UV emission is supposed to be mainly from the host galaxy and 
other sources rather than this synchrotron emission, in case of \fermi\ it is particularly dominated by an unknown 
continuum component that varies independent from the X\&$\gamma$-ray emission; therefore, the observed UV emission can 
only serve as an upper limit in our spectrum fitting, and a very hard electron spectral index of $-$1.4 is introduced 
in both the high state and the low state fittings. Noticeably, the intrinsic UV flux accompanied by the 
$\gamma-$ray flare could be much higher than this limit, due to the unknown extinction.

With such a hard electron spectrum, the inverse Compton (IC) fitting of the low state requires an electron 
cutoff energy of a quite low energy (8~GeV in Fig~\ref{sed}), in order to constrain the IC peak at $<$10$^{25}$~Hz; 
As a consequence of choosing such a low cutoff energy, a very compact gamma-ray emitting region is needed to boost 
up the IC flux. In Fig.~\ref{sed}, we adopt a $R_{\rm b}$ size of merely 2.3$\times$10$^{14}$ cm in the comoving frame, 
which corresponds to a minimum GeV variability of 10 minutes when using a Doppler factor of 25 
\citep[which is in the range of Doppler factors for blazars, e.g.,][]{Savolainen10,Fan14,Liodakis17}. 
As a comparison, the radius of the Innermost Stable Circular Orbit of a $3\times10^{8}M_\odot$ black hole 
\citep{Bruni18} is roughly 2.4$\times$10$^{14}$~cm.

The observed GeV spectrum of the high state is very hard and it allows the electron cutoff energy to move 
freely above 10~GeV during the fitting. This large parameter space of fitting the high state is also due to 
the lack of direct observational constraints on the magnetic field $B$, Doppler factor $\delta$, and the 
size of the gamma-ray emitting region $R_{\rm b}$. In Fig.~\ref{sed}, we have shown one of the many fitting 
results with an electron cutoff energy of 25~GeV. More parameter details about our fitting can be found in 
Table.~\ref{tab:sed_fit_par}.

\begin{table*}
\begin{center}
\caption{Parameters used in the SED fits. Here $E_{\rm e}$ only represents the total electron energy currently in the GeV emitting region, it is a free parameter, meanwhile the total energy output per second for a jet is around 10$^{45}$ erg/s.
\label{tab:sed_fit_par}}
\begin{tabular}{ccccccc}
\hline
\hline
Epoch & $B$ & $R_{\rm b}$ & $\delta$ & $p$ & $E_{\rm cut}$ & $E_{\rm e}$ \\
           & (G) &  (cm) & & & (GeV) & (erg) \\
\hline
May 2017 &0.4  & 3$\times$10$^{15}$ &   25   &  $-$1.4 &   25 &  2.16$\times$10$^{45}$ \\
July 2017 & 3    & 3.2$\times$10$^{14}$ & 25   &  $-$1.5 & 8 & 4.76$\times$10$^{43}$ \\
\hline
\end{tabular}
\end{center}
\end{table*}

\subsection{The mysterious optical variation and MIR flare}
\label{sub_sect:opt}
The strong optical variation of \fermi, as seen in Fig.~\ref{fig:mwl_lc}, does not show one-to-one 
consistency with the X-/\gr variation. Thus, besides the X/\gr\ flare component and the host 
galaxy, alternative sources are likely to dominate the optical band, e.g., other blobs in the jet or even from the core region. 

\gr\ sources in BL Lac objects are normally considered as the relativistic shocks inside the jet. 
It is well accepted that the jet flow is very likely to be intermittent \citep{wang09}. When an injected flow catches up with a 
slower flow or hits some over-dense medium, a shock is formed. The observed blobs in the jet, as seen in those best 
observed relativistic jets, e.g., M87 \citep{Owen1989} and 3C~120 \citep{Casadio2015}, are commonly interpreted as 
shock waves moving along the jet. In case of an intermittent power injection, the shock could slow down and the 
opening angle of the beaming emission will be widening, the rise and the decay of an internal shock could naturally 
cause a variety of light curves based on different viewing angles. Clearly, each shock (knot) of the jet could result 
different non-thermal emissions, see e.g. the well observed jet of 3C~273 and M87 by HST \citep{HST_M87,HST_3C273} 
and Chandra \citep{Chandra_M87,Chandra_3C273}, which has shown clearly resolved knots along the jet, and their peak 
energies gradually move from X-ray to optical with increasing distances to the nucleus. 

To explain the violent variation of this unknown optical source (with timescale down to 1 week) of \fermi, an intrinsic 
rise \& decay of the accelerator close to the BH could cause the optical flare. Additionally, many magnetic launching models 
suggests a jet field of helix structure, which is also been supported by the polarization observations \citep[e.g.,][]{Gabuzda2004}. 
In the helical model, strongest emission is obtained when the shock wave reaches the bent regions towards the observer \citep{Gomez1994}. 
In the case where an alternative blob dominating the optical flare through its synchrotron emission, correlated infrared flares 
are likely to be observed (which is indeed observed in 2018, Fig.~\ref{fig:oirlc}). 

In our simple SSC model above, the optical observation only functions as an upper-limit. Clearly, one-zone jet models, 
\citep[see, e.g.,][]{Bruni18}, which includes a relativistic jet with a single emitting zone, an accretion disk, and the 
host galaxy, will face difficulties to explain the observed UV excess, the strong optical variation and MIR flare seen in 2018.

\section{Conclusion and Outlook}

The high-energy transient \fermi\ is likely due to a sudden energy release (geometrical beaming may play a role as well) 
of a previously unknown BL Lac object, as first suggested by \citet{Bruni18} -- no high-energy emission was ever seen over 
the $\sim$9 years' lifetime of the {\it Fermi} satellite (nor by {\it MAXI} in X-rays) before April 2017. It is important 
to understand the mechanism that causes the sudden increase of radiation recently. We argue that a shock-in-jet model combined 
with viewing angle effects may explain the high-energy flares. 

\fermi\ displays the typical blazar characteristics including strong X/\gr\ flares after a decade-long quiescent period, a typical blazar SED, and rapid variations with timescale down to $<$1~hour, The optical flux of \fermi\ does not vary at the same time with the X/\gr\ flux. Thus, besides the X/\gr\ flare component and the host galaxy, alternative sources are likely to dominate the optical band.

Being a HSP BL Lac object, \fermi\ is likely a TeV-emitting blazar, and it shows huge variation in X-rays and \grs. Observing \fermi\ by 
current and/or upcoming {\it Cherenkov Telescope Array}, e.g., {\it CTA} will tell us about the position of the Compton peak, 
that in turn will constrain the jet physics. At times being an extreme blazar, it would also be another blazar to probe extreme 
particle acceleration, EBL, and/or EGMF.

\section{Acknowledgments}
We thank Nidia Morrell for helping during the Magellan observation, and Da-Hai Yan for useful discussion. 
PHT is supported by the National Science Foundation of China (NSFC) grants 11633007, 11661161010, and U1731136. 
JM is supported by the NSFC grants 11673062, the Hundred Talent Program of Chinese Academy of Sciences, the 
Major Program of Chinese Academy of Sciences (KJZD-EW-M06), and the Oversea Talent Program of Yunnan Province. 
JHF is supported by NSFC grants 11733001 and U1531245. This work is sponsored (in part) by the Chinese Academy of Sciences (CAS), 
through a grant to the CAS South America Center for Astronomy (CASSACA) in Santiago, Chile. This research made use of data supplied 
by the High Energy Astrophysics Science Archive Research Center (HEASARC) at NASA's Goddard Space Flight Center, and 
the UK Swift Science Data Centre at the University of Leicester. This publication makes use of data products from the 
Wide-field Infrared Survey Explorer (WISE), which is a joint project of the UCLA, and JPL/California Institute of Technology, 
funded by NASA. This publication also makes use of data products from NEOWISE, which is a project of the JPL/California Institute 
of Technology, funded by the Planetary Science Division of NASA. This work is based on observations obtained with XMM-Newton, 
an ESA science mission with instruments and contributions directly funded by ESA Member States and NASA. 
This paper includes data gathered with the 6.5 meter Magellan Telescopes located at Las Campanas Observatory, Chile.

\bibliographystyle{elsarticle-harv} 
\bibliography{j1544}

\begin{thebibliography}{61}
\expandafter\ifx\csname natexlab\endcsname\relax\def\natexlab#1{#1}\fi
\providecommand{\url}[1]{\texttt{#1}}
\providecommand{\href}[2]{#2}
\providecommand{\path}[1]{#1}
\providecommand{\DOIprefix}{doi:}
\providecommand{\ArXivprefix}{arXiv:}
\providecommand{\URLprefix}{URL: }
\providecommand{\Pubmedprefix}{pmid:}
\providecommand{\doi}[1]{\href{http://dx.doi.org/#1}{\path{#1}}}
\providecommand{\Pubmed}[1]{\href{pmid:#1}{\path{#1}}}
\providecommand{\bibinfo}[2]{#2}
\ifx\xfnm\relax \def\xfnm[#1]{\unskip,\space#1}\fi
\bibitem[{{Abdo} et~al.(2010){Abdo}, {Ackermann}, {Agudo}, {Ajello}, {Aller},
  {Aller}, {Angelakis}, {Arkharov}, {Axelsson} and {Bach}}]{Abdo10}
\bibinfo{author}{{Abdo}, A.A.}, \bibinfo{author}{{Ackermann}, M.},
  \bibinfo{author}{{Agudo}, I.}, \bibinfo{author}{{Ajello}, M.},
  \bibinfo{author}{{Aller}, H.D.}, \bibinfo{author}{{Aller}, M.F.},
  \bibinfo{author}{{Angelakis}, E.}, \bibinfo{author}{{Arkharov}, A.A.},
  \bibinfo{author}{{Axelsson}, M.}, \bibinfo{author}{{Bach}, U.},
  \bibinfo{year}{2010}.
\newblock \bibinfo{title}{{The Spectral Energy Distribution of Fermi Bright
  Blazars}}.
\newblock \bibinfo{journal}{\apj} \bibinfo{volume}{716},
  \bibinfo{pages}{30--70}.
\newblock \DOIprefix\doi{10.1088/0004-637X/716/1/30},
  \href{http://arxiv.org/abs/0912.2040}{{\tt arXiv:0912.2040}}.
\bibitem[{{Acero} et~al.(2015){Acero}, {Ackermann}, {Ajello}, {Albert},
  {Atwood}, {Axelsson}, {Baldini}, {Ballet}, {Barbiellini} and
  {Bastieri}}]{lat_3rd_cat}
\bibinfo{author}{{Acero}, F.}, \bibinfo{author}{{Ackermann}, M.},
  \bibinfo{author}{{Ajello}, M.}, \bibinfo{author}{{Albert}, A.},
  \bibinfo{author}{{Atwood}, W.B.}, \bibinfo{author}{{Axelsson}, M.},
  \bibinfo{author}{{Baldini}, L.}, \bibinfo{author}{{Ballet}, J.},
  \bibinfo{author}{{Barbiellini}, G.}, \bibinfo{author}{{Bastieri}, D.},
  \bibinfo{year}{2015}.
\newblock \bibinfo{title}{{Fermi Large Area Telescope Third Source Catalog}}.
\newblock \bibinfo{journal}{\apjs} \bibinfo{volume}{218}, \bibinfo{pages}{23}.
\newblock \DOIprefix\doi{10.1088/0067-0049/218/2/23},
  \href{http://arxiv.org/abs/1501.02003}{{\tt arXiv:1501.02003}}.
\bibitem[{{Ackermann} et~al.(2011){Ackermann}, {Ajello}, {Allafort},
  {Antolini}, {Atwood}, {Axelsson}, {Baldini}, {Ballet}, {Barbiellini} and
  {Bastieri}}]{second_lac}
\bibinfo{author}{{Ackermann}, M.}, \bibinfo{author}{{Ajello}, M.},
  \bibinfo{author}{{Allafort}, A.}, \bibinfo{author}{{Antolini}, E.},
  \bibinfo{author}{{Atwood}, W.B.}, \bibinfo{author}{{Axelsson}, M.},
  \bibinfo{author}{{Baldini}, L.}, \bibinfo{author}{{Ballet}, J.},
  \bibinfo{author}{{Barbiellini}, G.}, \bibinfo{author}{{Bastieri}, D.},
  \bibinfo{year}{2011}.
\newblock \bibinfo{title}{{The Second Catalog of Active Galactic Nuclei
  Detected by the Fermi Large Area Telescope}}.
\newblock \bibinfo{journal}{\apj} \bibinfo{volume}{743}, \bibinfo{pages}{171}.
\newblock \DOIprefix\doi{10.1088/0004-637X/743/2/171},
  \href{http://arxiv.org/abs/1108.1420}{{\tt arXiv:1108.1420}}.
\bibitem[{{Ahnen} et~al.(2018){Ahnen}, {Ansoldi}, {Antonelli}, {Arcaro},
  {Babi{\'c}}, {Banerjee}, {Bangale}, {Barres de Almeida}, {Barrio} and
  {Becerra Gonz{\'a}lez}}]{ahnen18}
\bibinfo{author}{{Ahnen}, M.L.}, \bibinfo{author}{{Ansoldi}, S.},
  \bibinfo{author}{{Antonelli}, L.A.}, \bibinfo{author}{{Arcaro}, C.},
  \bibinfo{author}{{Babi{\'c}}, A.}, \bibinfo{author}{{Banerjee}, B.},
  \bibinfo{author}{{Bangale}, P.}, \bibinfo{author}{{Barres de Almeida}, U.},
  \bibinfo{author}{{Barrio}, J.A.}, \bibinfo{author}{{Becerra Gonz{\'a}lez},
  J.}, \bibinfo{year}{2018}.
\newblock \bibinfo{title}{{Extreme HBL behavior of Markarian 501 during 2012}}.
\newblock \bibinfo{journal}{\aap} \bibinfo{volume}{620}, \bibinfo{pages}{A181}.
\newblock \DOIprefix\doi{10.1051/0004-6361/201833704},
  \href{http://arxiv.org/abs/1808.04300}{{\tt arXiv:1808.04300}}.
\bibitem[{{Alexander}(1997)}]{zdcf}
\bibinfo{author}{{Alexander}, T.}, \bibinfo{year}{1997}.
\newblock \bibinfo{title}{{Is AGN Variability Correlated with Other AGN
  Properties? ZDCF Analysis of Small Samples of Sparse Light Curves}}, in:
  \bibinfo{editor}{{Maoz}, D.}, \bibinfo{editor}{{Sternberg}, A.},
  \bibinfo{editor}{{Leibowitz}, E.M.} (Eds.), \bibinfo{booktitle}{Astronomical
  Time Series}, p. \bibinfo{pages}{163}.
\newblock \DOIprefix\doi{10.1007/978-94-015-8941-3_14}.
\bibitem[{{Astropy Collaboration} et~al.(2018){Astropy Collaboration},
  {Price-Whelan}, {Sip{\H{o}}cz}, {G{\"u}nther}, {Lim}, {Crawford}, {Conseil},
  {Shupe}, {Craig}, {Dencheva}, {Ginsburg}, {Vand erPlas}, {Bradley},
  {P{\'e}rez-Su{\'a}rez}, {de Val-Borro}, {Aldcroft}, {Cruz}, {Robitaille},
  {Tollerud}, {Ardelean}, {Babej}, {Bach}, {Bachetti}, {Bakanov}, {Bamford},
  {Barentsen}, {Barmby}, {Baumbach}, {Berry}, {Biscani}, {Boquien}, {Bostroem},
  {Bouma}, {Brammer}, {Bray}, {Breytenbach}, {Buddelmeijer}, {Burke},
  {Calderone}, {Cano Rodr{\'\i}guez}, {Cara}, {Cardoso}, {Cheedella}, {Copin},
  {Corrales}, {Crichton}, {D'Avella}, {Deil}, {Depagne}, {Dietrich}, {Donath},
  {Droettboom}, {Earl}, {Erben}, {Fabbro}, {Ferreira}, {Finethy}, {Fox},
  {Garrison}, {Gibbons}, {Goldstein}, {Gommers}, {Greco}, {Greenfield},
  {Groener}, {Grollier}, {Hagen}, {Hirst}, {Homeier}, {Horton}, {Hosseinzadeh},
  {Hu}, {Hunkeler}, {Ivezi{\'c}}, {Jain}, {Jenness}, {Kanarek}, {Kendrew},
  {Kern}, {Kerzendorf}, {Khvalko}, {King}, {Kirkby}, {Kulkarni}, {Kumar},
  {Lee}, {Lenz}, {Littlefair}, {Ma}, {Macleod}, {Mastropietro}, {McCully},
  {Montagnac}, {Morris}, {Mueller}, {Mumford}, {Muna}, {Murphy}, {Nelson},
  {Nguyen}, {Ninan}, {N{\"o}the}, {Ogaz}, {Oh}, {Parejko}, {Parley}, {Pascual},
  {Patil}, {Patil}, {Plunkett}, {Prochaska}, {Rastogi}, {Reddy Janga},
  {Sabater}, {Sakurikar}, {Seifert}, {Sherbert}, {Sherwood-Taylor}, {Shih},
  {Sick}, {Silbiger}, {Singanamalla}, {Singer}, {Sladen}, {Sooley},
  {Sornarajah}, {Streicher}, {Teuben}, {Thomas}, {Tremblay}, {Turner},
  {Terr{\'o}n}, {van Kerkwijk}, {de la Vega}, {Watkins}, {Weaver}, {Whitmore},
  {Woillez}, {Zabalza} and {Astropy Contributors}}]{astropy:2018}
\bibinfo{author}{{Astropy Collaboration}}, \bibinfo{author}{{Price-Whelan},
  A.M.}, \bibinfo{author}{{Sip{\H{o}}cz}, B.M.},
  \bibinfo{author}{{G{\"u}nther}, H.M.}, \bibinfo{author}{{Lim}, P.L.},
  \bibinfo{author}{{Crawford}, S.M.}, \bibinfo{author}{{Conseil}, S.},
  \bibinfo{author}{{Shupe}, D.L.}, \bibinfo{author}{{Craig}, M.W.},
  \bibinfo{author}{{Dencheva}, N.}, \bibinfo{author}{{Ginsburg}, A.},
  \bibinfo{author}{{Vand erPlas}, J.T.}, \bibinfo{author}{{Bradley}, L.D.},
  \bibinfo{author}{{P{\'e}rez-Su{\'a}rez}, D.}, \bibinfo{author}{{de
  Val-Borro}, M.}, \bibinfo{author}{{Aldcroft}, T.L.}, \bibinfo{author}{{Cruz},
  K.L.}, \bibinfo{author}{{Robitaille}, T.P.}, \bibinfo{author}{{Tollerud},
  E.J.}, \bibinfo{author}{{Ardelean}, C.}, \bibinfo{author}{{Babej}, T.},
  \bibinfo{author}{{Bach}, Y.P.}, \bibinfo{author}{{Bachetti}, M.},
  \bibinfo{author}{{Bakanov}, A.V.}, \bibinfo{author}{{Bamford}, S.P.},
  \bibinfo{author}{{Barentsen}, G.}, \bibinfo{author}{{Barmby}, P.},
  \bibinfo{author}{{Baumbach}, A.}, \bibinfo{author}{{Berry}, K.L.},
  \bibinfo{author}{{Biscani}, F.}, \bibinfo{author}{{Boquien}, M.},
  \bibinfo{author}{{Bostroem}, K.A.}, \bibinfo{author}{{Bouma}, L.G.},
  \bibinfo{author}{{Brammer}, G.B.}, \bibinfo{author}{{Bray}, E.M.},
  \bibinfo{author}{{Breytenbach}, H.}, \bibinfo{author}{{Buddelmeijer}, H.},
  \bibinfo{author}{{Burke}, D.J.}, \bibinfo{author}{{Calderone}, G.},
  \bibinfo{author}{{Cano Rodr{\'\i}guez}, J.L.}, \bibinfo{author}{{Cara}, M.},
  \bibinfo{author}{{Cardoso}, J.V.M.}, \bibinfo{author}{{Cheedella}, S.},
  \bibinfo{author}{{Copin}, Y.}, \bibinfo{author}{{Corrales}, L.},
  \bibinfo{author}{{Crichton}, D.}, \bibinfo{author}{{D'Avella}, D.},
  \bibinfo{author}{{Deil}, C.}, \bibinfo{author}{{Depagne}, {\'E}.},
  \bibinfo{author}{{Dietrich}, J.P.}, \bibinfo{author}{{Donath}, A.},
  \bibinfo{author}{{Droettboom}, M.}, \bibinfo{author}{{Earl}, N.},
  \bibinfo{author}{{Erben}, T.}, \bibinfo{author}{{Fabbro}, S.},
  \bibinfo{author}{{Ferreira}, L.A.}, \bibinfo{author}{{Finethy}, T.},
  \bibinfo{author}{{Fox}, R.T.}, \bibinfo{author}{{Garrison}, L.H.},
  \bibinfo{author}{{Gibbons}, S.L.J.}, \bibinfo{author}{{Goldstein}, D.A.},
  \bibinfo{author}{{Gommers}, R.}, \bibinfo{author}{{Greco}, J.P.},
  \bibinfo{author}{{Greenfield}, P.}, \bibinfo{author}{{Groener}, A.M.},
  \bibinfo{author}{{Grollier}, F.}, \bibinfo{author}{{Hagen}, A.},
  \bibinfo{author}{{Hirst}, P.}, \bibinfo{author}{{Homeier}, D.},
  \bibinfo{author}{{Horton}, A.J.}, \bibinfo{author}{{Hosseinzadeh}, G.},
  \bibinfo{author}{{Hu}, L.}, \bibinfo{author}{{Hunkeler}, J.S.},
  \bibinfo{author}{{Ivezi{\'c}}, {\v{Z}}.}, \bibinfo{author}{{Jain}, A.},
  \bibinfo{author}{{Jenness}, T.}, \bibinfo{author}{{Kanarek}, G.},
  \bibinfo{author}{{Kendrew}, S.}, \bibinfo{author}{{Kern}, N.S.},
  \bibinfo{author}{{Kerzendorf}, W.E.}, \bibinfo{author}{{Khvalko}, A.},
  \bibinfo{author}{{King}, J.}, \bibinfo{author}{{Kirkby}, D.},
  \bibinfo{author}{{Kulkarni}, A.M.}, \bibinfo{author}{{Kumar}, A.},
  \bibinfo{author}{{Lee}, A.}, \bibinfo{author}{{Lenz}, D.},
  \bibinfo{author}{{Littlefair}, S.P.}, \bibinfo{author}{{Ma}, Z.},
  \bibinfo{author}{{Macleod}, D.M.}, \bibinfo{author}{{Mastropietro}, M.},
  \bibinfo{author}{{McCully}, C.}, \bibinfo{author}{{Montagnac}, S.},
  \bibinfo{author}{{Morris}, B.M.}, \bibinfo{author}{{Mueller}, M.},
  \bibinfo{author}{{Mumford}, S.J.}, \bibinfo{author}{{Muna}, D.},
  \bibinfo{author}{{Murphy}, N.A.}, \bibinfo{author}{{Nelson}, S.},
  \bibinfo{author}{{Nguyen}, G.H.}, \bibinfo{author}{{Ninan}, J.P.},
  \bibinfo{author}{{N{\"o}the}, M.}, \bibinfo{author}{{Ogaz}, S.},
  \bibinfo{author}{{Oh}, S.}, \bibinfo{author}{{Parejko}, J.K.},
  \bibinfo{author}{{Parley}, N.}, \bibinfo{author}{{Pascual}, S.},
  \bibinfo{author}{{Patil}, R.}, \bibinfo{author}{{Patil}, A.A.},
  \bibinfo{author}{{Plunkett}, A.L.}, \bibinfo{author}{{Prochaska}, J.X.},
  \bibinfo{author}{{Rastogi}, T.}, \bibinfo{author}{{Reddy Janga}, V.},
  \bibinfo{author}{{Sabater}, J.}, \bibinfo{author}{{Sakurikar}, P.},
  \bibinfo{author}{{Seifert}, M.}, \bibinfo{author}{{Sherbert}, L.E.},
  \bibinfo{author}{{Sherwood-Taylor}, H.}, \bibinfo{author}{{Shih}, A.Y.},
  \bibinfo{author}{{Sick}, J.}, \bibinfo{author}{{Silbiger}, M.T.},
  \bibinfo{author}{{Singanamalla}, S.}, \bibinfo{author}{{Singer}, L.P.},
  \bibinfo{author}{{Sladen}, P.H.}, \bibinfo{author}{{Sooley}, K.A.},
  \bibinfo{author}{{Sornarajah}, S.}, \bibinfo{author}{{Streicher}, O.},
  \bibinfo{author}{{Teuben}, P.}, \bibinfo{author}{{Thomas}, S.W.},
  \bibinfo{author}{{Tremblay}, G.R.}, \bibinfo{author}{{Turner}, J.E.H.},
  \bibinfo{author}{{Terr{\'o}n}, V.}, \bibinfo{author}{{van Kerkwijk}, M.H.},
  \bibinfo{author}{{de la Vega}, A.}, \bibinfo{author}{{Watkins}, L.L.},
  \bibinfo{author}{{Weaver}, B.A.}, \bibinfo{author}{{Whitmore}, J.B.},
  \bibinfo{author}{{Woillez}, J.}, \bibinfo{author}{{Zabalza}, V.},
  \bibinfo{author}{{Astropy Contributors}}, \bibinfo{year}{2018}.
\newblock \bibinfo{title}{{The Astropy Project: Building an Open-science
  Project and Status of the v2.0 Core Package}}.
\newblock \bibinfo{journal}{\aj} \bibinfo{volume}{156}, \bibinfo{pages}{123}.
\newblock \DOIprefix\doi{10.3847/1538-3881/aabc4f},
  \href{http://arxiv.org/abs/1801.02634}{{\tt arXiv:1801.02634}}.
\bibitem[{{Astropy Collaboration} et~al.(2013){Astropy Collaboration},
  {Robitaille}, {Tollerud}, {Greenfield}, {Droettboom}, {Bray}, {Aldcroft},
  {Davis}, {Ginsburg}, {Price-Whelan}, {Kerzendorf}, {Conley}, {Crighton},
  {Barbary}, {Muna}, {Ferguson}, {Grollier}, {Parikh}, {Nair}, {Unther},
  {Deil}, {Woillez}, {Conseil}, {Kramer}, {Turner}, {Singer}, {Fox}, {Weaver},
  {Zabalza}, {Edwards}, {Azalee Bostroem}, {Burke}, {Casey}, {Crawford},
  {Dencheva}, {Ely}, {Jenness}, {Labrie}, {Lim}, {Pierfederici}, {Pontzen},
  {Ptak}, {Refsdal}, {Servillat} and {Streicher}}]{astropy:2013}
\bibinfo{author}{{Astropy Collaboration}}, \bibinfo{author}{{Robitaille},
  T.P.}, \bibinfo{author}{{Tollerud}, E.J.}, \bibinfo{author}{{Greenfield},
  P.}, \bibinfo{author}{{Droettboom}, M.}, \bibinfo{author}{{Bray}, E.},
  \bibinfo{author}{{Aldcroft}, T.}, \bibinfo{author}{{Davis}, M.},
  \bibinfo{author}{{Ginsburg}, A.}, \bibinfo{author}{{Price-Whelan}, A.M.},
  \bibinfo{author}{{Kerzendorf}, W.E.}, \bibinfo{author}{{Conley}, A.},
  \bibinfo{author}{{Crighton}, N.}, \bibinfo{author}{{Barbary}, K.},
  \bibinfo{author}{{Muna}, D.}, \bibinfo{author}{{Ferguson}, H.},
  \bibinfo{author}{{Grollier}, F.}, \bibinfo{author}{{Parikh}, M.M.},
  \bibinfo{author}{{Nair}, P.H.}, \bibinfo{author}{{Unther}, H.M.},
  \bibinfo{author}{{Deil}, C.}, \bibinfo{author}{{Woillez}, J.},
  \bibinfo{author}{{Conseil}, S.}, \bibinfo{author}{{Kramer}, R.},
  \bibinfo{author}{{Turner}, J.E.H.}, \bibinfo{author}{{Singer}, L.},
  \bibinfo{author}{{Fox}, R.}, \bibinfo{author}{{Weaver}, B.A.},
  \bibinfo{author}{{Zabalza}, V.}, \bibinfo{author}{{Edwards}, Z.I.},
  \bibinfo{author}{{Azalee Bostroem}, K.}, \bibinfo{author}{{Burke}, D.J.},
  \bibinfo{author}{{Casey}, A.R.}, \bibinfo{author}{{Crawford}, S.M.},
  \bibinfo{author}{{Dencheva}, N.}, \bibinfo{author}{{Ely}, J.},
  \bibinfo{author}{{Jenness}, T.}, \bibinfo{author}{{Labrie}, K.},
  \bibinfo{author}{{Lim}, P.L.}, \bibinfo{author}{{Pierfederici}, F.},
  \bibinfo{author}{{Pontzen}, A.}, \bibinfo{author}{{Ptak}, A.},
  \bibinfo{author}{{Refsdal}, B.}, \bibinfo{author}{{Servillat}, M.},
  \bibinfo{author}{{Streicher}, O.}, \bibinfo{year}{2013}.
\newblock \bibinfo{title}{{Astropy: A community Python package for astronomy}}.
\newblock \bibinfo{journal}{\aap} \bibinfo{volume}{558}, \bibinfo{pages}{A33}.
\newblock \DOIprefix\doi{10.1051/0004-6361/201322068},
  \href{http://arxiv.org/abs/1307.6212}{{\tt arXiv:1307.6212}}.
\bibitem[{{Atwood} et~al.(2009){Atwood}, {Abdo}, {Ackermann}, {Althouse},
  {Anderson}, {Axelsson}, {Baldini}, {Ballet}, {Band} and
  {Barbiellini}}]{lat_technical}
\bibinfo{author}{{Atwood}, W.B.}, \bibinfo{author}{{Abdo}, A.A.},
  \bibinfo{author}{{Ackermann}, M.}, \bibinfo{author}{{Althouse}, W.},
  \bibinfo{author}{{Anderson}, B.}, \bibinfo{author}{{Axelsson}, M.},
  \bibinfo{author}{{Baldini}, L.}, \bibinfo{author}{{Ballet}, J.},
  \bibinfo{author}{{Band}, D.L.}, \bibinfo{author}{{Barbiellini}, G.},
  \bibinfo{year}{2009}.
\newblock \bibinfo{title}{{The Large Area Telescope on the Fermi Gamma-Ray
  Space Telescope Mission}}.
\newblock \bibinfo{journal}{\apj} \bibinfo{volume}{697},
  \bibinfo{pages}{1071--1102}.
\newblock \DOIprefix\doi{10.1088/0004-637X/697/2/1071},
  \href{http://arxiv.org/abs/0902.1089}{{\tt arXiv:0902.1089}}.
\bibitem[{{Baars} et~al.(1977){Baars}, {Genzel}, {Pauliny-Toth} and
  {Witzel}}]{1977A&A....61...99B}
\bibinfo{author}{{Baars}, J.W.M.}, \bibinfo{author}{{Genzel}, R.},
  \bibinfo{author}{{Pauliny-Toth}, I.I.K.}, \bibinfo{author}{{Witzel}, A.},
  \bibinfo{year}{1977}.
\newblock \bibinfo{title}{{Reprint of 1977A\&amp;A....61...99B. The absolute
  spectrum of Cas A; an accurate flux density scale and a set of secondary
  calibrators.}}
\newblock \bibinfo{journal}{\aap} \bibinfo{volume}{500},
  \bibinfo{pages}{135--142}.
\bibitem[{{Bahcall} et~al.(1995){Bahcall}, {Kirhakos}, {Schneider}, {Davis},
  {Muxlow}, {Garrington}, {Conway} and {Unwin}}]{HST_3C273}
\bibinfo{author}{{Bahcall}, J.N.}, \bibinfo{author}{{Kirhakos}, S.},
  \bibinfo{author}{{Schneider}, D.P.}, \bibinfo{author}{{Davis}, R.J.},
  \bibinfo{author}{{Muxlow}, T.W.B.}, \bibinfo{author}{{Garrington}, S.T.},
  \bibinfo{author}{{Conway}, R.G.}, \bibinfo{author}{{Unwin}, S.C.},
  \bibinfo{year}{1995}.
\newblock \bibinfo{title}{{Hubble Space Telescope and MERLIN Observations of
  the Jet in 3C 273}}.
\newblock \bibinfo{journal}{\apj} \bibinfo{volume}{452}, \bibinfo{pages}{L91}.
\newblock \DOIprefix\doi{10.1086/309717},
  \href{http://arxiv.org/abs/astro-ph/9509028}{{\tt arXiv:astro-ph/9509028}}.
\bibitem[{{Biretta} et~al.(1999){Biretta}, {Sparks} and {Macchetto}}]{HST_M87}
\bibinfo{author}{{Biretta}, J.A.}, \bibinfo{author}{{Sparks}, W.B.},
  \bibinfo{author}{{Macchetto}, F.}, \bibinfo{year}{1999}.
\newblock \bibinfo{title}{{Hubble Space Telescope Observations of Superluminal
  Motion in the M87 Jet}}.
\newblock \bibinfo{journal}{\apj} \bibinfo{volume}{520},
  \bibinfo{pages}{621--626}.
\newblock \DOIprefix\doi{10.1086/307499}.
\bibitem[{{Boller} et~al.(2016){Boller}, {Freyberg}, {Tr{\"u}mper}, {Haberl},
  {Voges} and {Nandra}}]{2rxs}
\bibinfo{author}{{Boller}, T.}, \bibinfo{author}{{Freyberg}, M.J.},
  \bibinfo{author}{{Tr{\"u}mper}, J.}, \bibinfo{author}{{Haberl}, F.},
  \bibinfo{author}{{Voges}, W.}, \bibinfo{author}{{Nandra}, K.},
  \bibinfo{year}{2016}.
\newblock \bibinfo{title}{{Second ROSAT all-sky survey (2RXS) source
  catalogue}}.
\newblock \bibinfo{journal}{\aap} \bibinfo{volume}{588}, \bibinfo{pages}{A103}.
\newblock \DOIprefix\doi{10.1051/0004-6361/201525648},
  \href{http://arxiv.org/abs/1609.09244}{{\tt arXiv:1609.09244}}.
\bibitem[{Bowley(1928)}]{bowley28}
\bibinfo{author}{Bowley, A.L.}, \bibinfo{year}{1928}.
\newblock \bibinfo{title}{The standard deviation of the correlation
  coefficient}.
\newblock \bibinfo{journal}{Journal of the American Statistical Association}
  \bibinfo{volume}{23}, \bibinfo{pages}{31--34}.
\newblock \URLprefix \url{http://www.jstor.org/stable/2277400}.
\bibitem[{{Bruni} et~al.(2018){Bruni}, {Panessa}, {Ghisellini}, {Chavushyan},
  {Pe{\~n}a-Herazo}, {Hern{\'a}ndez-Garc{\'\i}a}, {Bazzano}, {Ubertini} and
  {Kraus}}]{Bruni18}
\bibinfo{author}{{Bruni}, G.}, \bibinfo{author}{{Panessa}, F.},
  \bibinfo{author}{{Ghisellini}, G.}, \bibinfo{author}{{Chavushyan}, V.},
  \bibinfo{author}{{Pe{\~n}a-Herazo}, H.A.},
  \bibinfo{author}{{Hern{\'a}ndez-Garc{\'\i}a}, L.},
  \bibinfo{author}{{Bazzano}, A.}, \bibinfo{author}{{Ubertini}, P.},
  \bibinfo{author}{{Kraus}, A.}, \bibinfo{year}{2018}.
\newblock \bibinfo{title}{{Fermi Transient J1544-0649: A Flaring Radio-weak BL
  Lac}}.
\newblock \bibinfo{journal}{\apj} \bibinfo{volume}{854}, \bibinfo{pages}{L23}.
\newblock \DOIprefix\doi{10.3847/2041-8213/aaacfb},
  \href{http://arxiv.org/abs/1802.01105}{{\tt arXiv:1802.01105}}.
\bibitem[{{Casadio} et~al.(2015){Casadio}, {G{\'o}mez}, {Grandi}, {Jorstad},
  {Marscher}, {Lister}, {Kovalev}, {Savolainen} and {Pushkarev}}]{Casadio2015}
\bibinfo{author}{{Casadio}, C.}, \bibinfo{author}{{G{\'o}mez}, J.L.},
  \bibinfo{author}{{Grandi}, P.}, \bibinfo{author}{{Jorstad}, S.G.},
  \bibinfo{author}{{Marscher}, A.P.}, \bibinfo{author}{{Lister}, M.L.},
  \bibinfo{author}{{Kovalev}, Y.Y.}, \bibinfo{author}{{Savolainen}, T.},
  \bibinfo{author}{{Pushkarev}, A.B.}, \bibinfo{year}{2015}.
\newblock \bibinfo{title}{{The Connection between the Radio Jet and the
  Gamma-ray Emission in the Radio Galaxy 3C 120}}.
\newblock \bibinfo{journal}{\apj} \bibinfo{volume}{808}, \bibinfo{pages}{162}.
\newblock \DOIprefix\doi{10.1088/0004-637X/808/2/162},
  \href{http://arxiv.org/abs/1505.03871}{{\tt arXiv:1505.03871}}.
\bibitem[{{Chornock} and {Margutti}(2017)}]{Chornock17}
\bibinfo{author}{{Chornock}, R.}, \bibinfo{author}{{Margutti}, R.},
  \bibinfo{year}{2017}.
\newblock \bibinfo{title}{{MDM Redshift of the Host of ASASSN-17gs}}.
\newblock \bibinfo{journal}{The Astronomer's Telegram} \bibinfo{volume}{10491},
  \bibinfo{pages}{1}.
\bibitem[{{Condon} et~al.(1998){Condon}, {Cotton}, {Greisen}, {Yin}, {Perley},
  {Taylor} and {Broderick}}]{NVSS_survey}
\bibinfo{author}{{Condon}, J.J.}, \bibinfo{author}{{Cotton}, W.D.},
  \bibinfo{author}{{Greisen}, E.W.}, \bibinfo{author}{{Yin}, Q.F.},
  \bibinfo{author}{{Perley}, R.A.}, \bibinfo{author}{{Taylor}, G.B.},
  \bibinfo{author}{{Broderick}, J.J.}, \bibinfo{year}{1998}.
\newblock \bibinfo{title}{{The NRAO VLA Sky Survey}}.
\newblock \bibinfo{journal}{\aj} \bibinfo{volume}{115},
  \bibinfo{pages}{1693--1716}.
\newblock \DOIprefix\doi{10.1086/300337}.
\bibitem[{{Costamante} et~al.(2018){Costamante}, {Bonnoli}, {Tavecchio},
  {Ghisellini}, {Tagliaferri} and {Khangulyan}}]{Costamante18}
\bibinfo{author}{{Costamante}, L.}, \bibinfo{author}{{Bonnoli}, G.},
  \bibinfo{author}{{Tavecchio}, F.}, \bibinfo{author}{{Ghisellini}, G.},
  \bibinfo{author}{{Tagliaferri}, G.}, \bibinfo{author}{{Khangulyan}, D.},
  \bibinfo{year}{2018}.
\newblock \bibinfo{title}{{The NuSTAR view on hard-TeV BL Lacs}}.
\newblock \bibinfo{journal}{\mnras} \bibinfo{volume}{477},
  \bibinfo{pages}{4257--4268}.
\newblock \DOIprefix\doi{10.1093/mnras/sty857},
  \href{http://arxiv.org/abs/1711.06282}{{\tt arXiv:1711.06282}}.
\bibitem[{{Costamante} et~al.(2001){Costamante}, {Ghisellini}, {Giommi},
  {Tagliaferri}, {Celotti}, {Chiaberge}, {Fossati}, {Maraschi}, {Tavecchio} and
  {Treves}}]{Costamante01}
\bibinfo{author}{{Costamante}, L.}, \bibinfo{author}{{Ghisellini}, G.},
  \bibinfo{author}{{Giommi}, P.}, \bibinfo{author}{{Tagliaferri}, G.},
  \bibinfo{author}{{Celotti}, A.}, \bibinfo{author}{{Chiaberge}, M.},
  \bibinfo{author}{{Fossati}, G.}, \bibinfo{author}{{Maraschi}, L.},
  \bibinfo{author}{{Tavecchio}, F.}, \bibinfo{author}{{Treves}, A.},
  \bibinfo{year}{2001}.
\newblock \bibinfo{title}{{Extreme synchrotron BL Lac objects. Stretching the
  blazar sequence}}.
\newblock \bibinfo{journal}{\aap} \bibinfo{volume}{371},
  \bibinfo{pages}{512--526}.
\newblock \DOIprefix\doi{10.1051/0004-6361:20010412},
  \href{http://arxiv.org/abs/astro-ph/0103343}{{\tt arXiv:astro-ph/0103343}}.
\bibitem[{{Dou} et~al.(2016){Dou}, {Wang}, {Jiang}, {Yang}, {Lyu} and
  {Zhou}}]{Dou2016}
\bibinfo{author}{{Dou}, L.}, \bibinfo{author}{{Wang}, T.g.},
  \bibinfo{author}{{Jiang}, N.}, \bibinfo{author}{{Yang}, C.},
  \bibinfo{author}{{Lyu}, J.}, \bibinfo{author}{{Zhou}, H.},
  \bibinfo{year}{2016}.
\newblock \bibinfo{title}{{Long Fading Mid-infrared Emission in Transient
  Coronal Line Emitters: Dust Echo of a Tidal Disruption Flare}}.
\newblock \bibinfo{journal}{\apj} \bibinfo{volume}{832}, \bibinfo{pages}{188}.
\newblock \DOIprefix\doi{10.3847/0004-637X/832/2/188},
  \href{http://arxiv.org/abs/1605.05145}{{\tt arXiv:1605.05145}}.
\bibitem[{{Drake} et~al.(2009){Drake}, {Djorgovski}, {Mahabal}, {Beshore},
  {Larson}, {Graham}, {Williams}, {Christensen}, {Catelan} and
  {Boattini}}]{Drake09}
\bibinfo{author}{{Drake}, A.J.}, \bibinfo{author}{{Djorgovski}, S.G.},
  \bibinfo{author}{{Mahabal}, A.}, \bibinfo{author}{{Beshore}, E.},
  \bibinfo{author}{{Larson}, S.}, \bibinfo{author}{{Graham}, M.J.},
  \bibinfo{author}{{Williams}, R.}, \bibinfo{author}{{Christensen}, E.},
  \bibinfo{author}{{Catelan}, M.}, \bibinfo{author}{{Boattini}, A.},
  \bibinfo{year}{2009}.
\newblock \bibinfo{title}{{First Results from the Catalina Real-Time Transient
  Survey}}.
\newblock \bibinfo{journal}{\apj} \bibinfo{volume}{696},
  \bibinfo{pages}{870--884}.
\newblock \DOIprefix\doi{10.1088/0004-637X/696/1/870},
  \href{http://arxiv.org/abs/0809.1394}{{\tt arXiv:0809.1394}}.
\bibitem[{{Fan} et~al.(2014){Fan}, {Bastieri}, {Yang}, {Liu}, {Hua}, {Yuan} and
  {Wu}}]{Fan14}
\bibinfo{author}{{Fan}, J.H.}, \bibinfo{author}{{Bastieri}, D.},
  \bibinfo{author}{{Yang}, J.H.}, \bibinfo{author}{{Liu}, Y.},
  \bibinfo{author}{{Hua}, T.X.}, \bibinfo{author}{{Yuan}, Y.H.},
  \bibinfo{author}{{Wu}, D.X.}, \bibinfo{year}{2014}.
\newblock \bibinfo{title}{{The lower limit of the Doppler factor for a Fermi
  blazar sample}}.
\newblock \bibinfo{journal}{Research in Astronomy and Astrophysics}
  \bibinfo{volume}{14}, \bibinfo{pages}{1135--1145}.
\newblock \DOIprefix\doi{10.1088/1674-4527/14/9/004}.
\bibitem[{{Fan} et~al.(2016){Fan}, {Yang}, {Liu}, {Luo}, {Lin}, {Yuan}, {Xiao},
  {Zhou}, {Hua} and {Pei}}]{Fan16}
\bibinfo{author}{{Fan}, J.H.}, \bibinfo{author}{{Yang}, J.H.},
  \bibinfo{author}{{Liu}, Y.}, \bibinfo{author}{{Luo}, G.Y.},
  \bibinfo{author}{{Lin}, C.}, \bibinfo{author}{{Yuan}, Y.H.},
  \bibinfo{author}{{Xiao}, H.B.}, \bibinfo{author}{{Zhou}, A.Y.},
  \bibinfo{author}{{Hua}, T.X.}, \bibinfo{author}{{Pei}, Z.Y.},
  \bibinfo{year}{2016}.
\newblock \bibinfo{title}{{The Spectral Energy Distributions of Fermi
  Blazars}}.
\newblock \bibinfo{journal}{\apjs} \bibinfo{volume}{226}, \bibinfo{pages}{20}.
\newblock \DOIprefix\doi{10.3847/0067-0049/226/2/20},
  \href{http://arxiv.org/abs/1608.03958}{{\tt arXiv:1608.03958}}.
\bibitem[{{Gabuzda} et~al.(2004){Gabuzda}, {Murray} and {Cronin}}]{Gabuzda2004}
\bibinfo{author}{{Gabuzda}, D.C.}, \bibinfo{author}{{Murray}, {\'E}.},
  \bibinfo{author}{{Cronin}, P.}, \bibinfo{year}{2004}.
\newblock \bibinfo{title}{{Helical magnetic fields associated with the
  relativistic jets of four BL Lac objects}}.
\newblock \bibinfo{journal}{\mnras} \bibinfo{volume}{351},
  \bibinfo{pages}{L89--L93}.
\newblock \DOIprefix\doi{10.1111/j.1365-2966.2004.08037.x},
  \href{http://arxiv.org/abs/astro-ph/0405394}{{\tt arXiv:astro-ph/0405394}}.
\bibitem[{{Gomez} et~al.(1994){Gomez}, {Alberdi} and {Marcaide}}]{Gomez1994}
\bibinfo{author}{{Gomez}, J.L.}, \bibinfo{author}{{Alberdi}, A.},
  \bibinfo{author}{{Marcaide}, J.M.}, \bibinfo{year}{1994}.
\newblock \bibinfo{title}{{Synchrotron emission from bent shocked relativistic
  jets. II. Shock waves in helical jets.}}
\newblock \bibinfo{journal}{\aap} \bibinfo{volume}{284},
  \bibinfo{pages}{51--64}.
\bibitem[{{Heckman} and {Best}(2014)}]{2014ARA&A..52..589H}
\bibinfo{author}{{Heckman}, T.M.}, \bibinfo{author}{{Best}, P.N.},
  \bibinfo{year}{2014}.
\newblock \bibinfo{title}{{The Coevolution of Galaxies and Supermassive Black
  Holes: Insights from Surveys of the Contemporary Universe}}.
\newblock \bibinfo{journal}{\araa} \bibinfo{volume}{52},
  \bibinfo{pages}{589--660}.
\newblock \DOIprefix\doi{10.1146/annurev-astro-081913-035722},
  \href{http://arxiv.org/abs/1403.4620}{{\tt arXiv:1403.4620}}.
\bibitem[{{Intema} et~al.(2017){Intema}, {Jagannathan}, {Mooley} and
  {Frail}}]{GMRT_survey}
\bibinfo{author}{{Intema}, H.T.}, \bibinfo{author}{{Jagannathan}, P.},
  \bibinfo{author}{{Mooley}, K.P.}, \bibinfo{author}{{Frail}, D.A.},
  \bibinfo{year}{2017}.
\newblock \bibinfo{title}{{The GMRT 150 MHz all-sky radio survey. First
  alternative data release TGSS ADR1}}.
\newblock \bibinfo{journal}{\aap} \bibinfo{volume}{598}, \bibinfo{pages}{A78}.
\newblock \DOIprefix\doi{10.1051/0004-6361/201628536},
  \href{http://arxiv.org/abs/1603.04368}{{\tt arXiv:1603.04368}}.
\bibitem[{{Jiang}(2018)}]{Jiang18}
\bibinfo{author}{{Jiang}, N.}, \bibinfo{year}{2018}.
\newblock \bibinfo{title}{{Intraday Mid-infrared Variability of CTA 102 During
  Its 2016 Giant Outburst}}.
\newblock \bibinfo{journal}{Research Notes of the American Astronomical
  Society} \bibinfo{volume}{2}, \bibinfo{pages}{134}.
\newblock \DOIprefix\doi{10.3847/2515-5172/aad693}.
\bibitem[{{Jiang} et~al.(2016){Jiang}, {Dou}, {Wang}, {Yang}, {Lyu} and
  {Zhou}}]{Jiang16}
\bibinfo{author}{{Jiang}, N.}, \bibinfo{author}{{Dou}, L.},
  \bibinfo{author}{{Wang}, T.}, \bibinfo{author}{{Yang}, C.},
  \bibinfo{author}{{Lyu}, J.}, \bibinfo{author}{{Zhou}, H.},
  \bibinfo{year}{2016}.
\newblock \bibinfo{title}{{The WISE Detection of an Infrared Echo in Tidal
  Disruption Event ASASSN-14li}}.
\newblock \bibinfo{journal}{\apj} \bibinfo{volume}{828}, \bibinfo{pages}{L14}.
\newblock \DOIprefix\doi{10.3847/2041-8205/828/1/L14},
  \href{http://arxiv.org/abs/1605.04640}{{\tt arXiv:1605.04640}}.
\bibitem[{{Jiang} et~al.(2012){Jiang}, {Zhou}, {Ho}, {Yuan}, {Wang}, {Dong},
  {Jiang}, {Ji} and {Tian}}]{Jiang12}
\bibinfo{author}{{Jiang}, N.}, \bibinfo{author}{{Zhou}, H.Y.},
  \bibinfo{author}{{Ho}, L.C.}, \bibinfo{author}{{Yuan}, W.},
  \bibinfo{author}{{Wang}, T.G.}, \bibinfo{author}{{Dong}, X.B.},
  \bibinfo{author}{{Jiang}, P.}, \bibinfo{author}{{Ji}, T.},
  \bibinfo{author}{{Tian}, Q.}, \bibinfo{year}{2012}.
\newblock \bibinfo{title}{{Rapid Infrared Variability of Three Radio-loud
  Narrow-line Seyfert 1 Galaxies: A View from the Wide-field Infrared Survey
  Explorer}}.
\newblock \bibinfo{journal}{\apj} \bibinfo{volume}{759}, \bibinfo{pages}{L31}.
\newblock \DOIprefix\doi{10.1088/2041-8205/759/2/L31},
  \href{http://arxiv.org/abs/1210.2800}{{\tt arXiv:1210.2800}}.
\bibitem[{{Kalberla} et~al.(2005){Kalberla}, {Burton}, {Hartmann}, {Arnal},
  {Bajaja}, {Morras} and {P{\"o}ppel}}]{Kalberla05}
\bibinfo{author}{{Kalberla}, P.M.W.}, \bibinfo{author}{{Burton}, W.B.},
  \bibinfo{author}{{Hartmann}, D.}, \bibinfo{author}{{Arnal}, E.M.},
  \bibinfo{author}{{Bajaja}, E.}, \bibinfo{author}{{Morras}, R.},
  \bibinfo{author}{{P{\"o}ppel}, W.G.L.}, \bibinfo{year}{2005}.
\newblock \bibinfo{title}{{The Leiden/Argentine/Bonn (LAB) Survey of Galactic
  HI. Final data release of the combined LDS and IAR surveys with improved
  stray-radiation corrections}}.
\newblock \bibinfo{journal}{\aap} \bibinfo{volume}{440},
  \bibinfo{pages}{775--782}.
\newblock \DOIprefix\doi{10.1051/0004-6361:20041864},
  \href{http://arxiv.org/abs/astro-ph/0504140}{{\tt arXiv:astro-ph/0504140}}.
\bibitem[{{Kochanek} et~al.(2017){Kochanek}, {Shappee}, {Stanek}, {Holoien},
  {Thompson}, {Prieto}, {Dong}, {Shields}, {Will} and {Britt}}]{Kochanek17}
\bibinfo{author}{{Kochanek}, C.S.}, \bibinfo{author}{{Shappee}, B.J.},
  \bibinfo{author}{{Stanek}, K.Z.}, \bibinfo{author}{{Holoien}, T.W.S.},
  \bibinfo{author}{{Thompson}, T.A.}, \bibinfo{author}{{Prieto}, J.L.},
  \bibinfo{author}{{Dong}, S.}, \bibinfo{author}{{Shields}, J.V.},
  \bibinfo{author}{{Will}, D.}, \bibinfo{author}{{Britt}, C.},
  \bibinfo{year}{2017}.
\newblock \bibinfo{title}{{The All-Sky Automated Survey for Supernovae
  (ASAS-SN) Light Curve Server v1.0}}.
\newblock \bibinfo{journal}{\pasp} \bibinfo{volume}{129},
  \bibinfo{pages}{104502}.
\newblock \DOIprefix\doi{10.1088/1538-3873/aa80d9},
  \href{http://arxiv.org/abs/1706.07060}{{\tt arXiv:1706.07060}}.
\bibitem[{{Liodakis} et~al.(2017){Liodakis}, {Marchili}, {Angelakis},
  {Fuhrmann}, {Nestoras}, {Myserlis}, {Karamanavis}, {Krichbaum}, {Sievers} and
  {Ungerechts}}]{Liodakis17}
\bibinfo{author}{{Liodakis}, I.}, \bibinfo{author}{{Marchili}, N.},
  \bibinfo{author}{{Angelakis}, E.}, \bibinfo{author}{{Fuhrmann}, L.},
  \bibinfo{author}{{Nestoras}, I.}, \bibinfo{author}{{Myserlis}, I.},
  \bibinfo{author}{{Karamanavis}, V.}, \bibinfo{author}{{Krichbaum}, T.P.},
  \bibinfo{author}{{Sievers}, A.}, \bibinfo{author}{{Ungerechts}, H.},
  \bibinfo{year}{2017}.
\newblock \bibinfo{title}{{F-GAMMA: variability Doppler factors of blazars from
  multiwavelength monitoring}}.
\newblock \bibinfo{journal}{\mnras} \bibinfo{volume}{466},
  \bibinfo{pages}{4625--4632}.
\newblock \DOIprefix\doi{10.1093/mnras/stx002},
  \href{http://arxiv.org/abs/1701.01452}{{\tt arXiv:1701.01452}}.
\bibitem[{{Lynden-Bell}(1969)}]{Lynden-Bell_69}
\bibinfo{author}{{Lynden-Bell}, D.}, \bibinfo{year}{1969}.
\newblock \bibinfo{title}{{Galactic Nuclei as Collapsed Old Quasars}}.
\newblock \bibinfo{journal}{\nat} \bibinfo{volume}{223},
  \bibinfo{pages}{690--694}.
\newblock \DOIprefix\doi{10.1038/223690a0}.
\bibitem[{{Mainzer} et~al.(2014){Mainzer}, {Bauer}, {Cutri}, {Grav}, {Masiero},
  {Beck}, {Clarkson}, {Conrow}, {Dailey} and {Eisenhardt}}]{Mainzer14}
\bibinfo{author}{{Mainzer}, A.}, \bibinfo{author}{{Bauer}, J.},
  \bibinfo{author}{{Cutri}, R.M.}, \bibinfo{author}{{Grav}, T.},
  \bibinfo{author}{{Masiero}, J.}, \bibinfo{author}{{Beck}, R.},
  \bibinfo{author}{{Clarkson}, P.}, \bibinfo{author}{{Conrow}, T.},
  \bibinfo{author}{{Dailey}, J.}, \bibinfo{author}{{Eisenhardt}, P.},
  \bibinfo{year}{2014}.
\newblock \bibinfo{title}{{Initial Performance of the NEOWISE Reactivation
  Mission}}.
\newblock \bibinfo{journal}{\apj} \bibinfo{volume}{792}, \bibinfo{pages}{30}.
\newblock \DOIprefix\doi{10.1088/0004-637X/792/1/30},
  \href{http://arxiv.org/abs/1406.6025}{{\tt arXiv:1406.6025}}.
\bibitem[{{Mingaliev} et~al.(2017){Mingaliev}, {Sotnikova}, {Mufakharov},
  {Nieppola}, {Tornikoski}, {Tammi}, {L{\"a}hteenm{\"a}ki}, {Udovitskiy} and
  {Erkenov}}]{2017AN....338..700M}
\bibinfo{author}{{Mingaliev}, M.}, \bibinfo{author}{{Sotnikova}, Y.},
  \bibinfo{author}{{Mufakharov}, T.}, \bibinfo{author}{{Nieppola}, E.},
  \bibinfo{author}{{Tornikoski}, M.}, \bibinfo{author}{{Tammi}, J.},
  \bibinfo{author}{{L{\"a}hteenm{\"a}ki}, A.}, \bibinfo{author}{{Udovitskiy},
  R.}, \bibinfo{author}{{Erkenov}, A.}, \bibinfo{year}{2017}.
\newblock \bibinfo{title}{{Simultaneous spectra and radio properties of BL
  Lacs}}.
\newblock \bibinfo{journal}{Astronomische Nachrichten} \bibinfo{volume}{338},
  \bibinfo{pages}{700--714}.
\newblock \DOIprefix\doi{10.1002/asna.201713361},
  \href{http://arxiv.org/abs/1707.07949}{{\tt arXiv:1707.07949}}.
\bibitem[{{Mingaliev} et~al.(2012){Mingaliev}, {Sotnikova}, {Torniainen},
  {Tornikoski} and {Udovitskiy}}]{2012A&A...544A..25M}
\bibinfo{author}{{Mingaliev}, M.G.}, \bibinfo{author}{{Sotnikova}, Y.V.},
  \bibinfo{author}{{Torniainen}, I.}, \bibinfo{author}{{Tornikoski}, M.},
  \bibinfo{author}{{Udovitskiy}, R.Y.}, \bibinfo{year}{2012}.
\newblock \bibinfo{title}{{Multifrequency study of GHz-peaked spectrum sources
  and candidates with the RATAN-600 radio telescope}}.
\newblock \bibinfo{journal}{\aap} \bibinfo{volume}{544}, \bibinfo{pages}{A25}.
\newblock \DOIprefix\doi{10.1051/0004-6361/201118506}.
\bibitem[{{Mingaliev} et~al.(2014){Mingaliev}, {Sotnikova}, {Udovitskiy},
  {Mufakharov}, {Nieppola} and {Erkenov}}]{2014arXiv1410.2835M}
\bibinfo{author}{{Mingaliev}, M.G.}, \bibinfo{author}{{Sotnikova}, Y.V.},
  \bibinfo{author}{{Udovitskiy}, R.Y.}, \bibinfo{author}{{Mufakharov}, T.V.},
  \bibinfo{author}{{Nieppola}, E.}, \bibinfo{author}{{Erkenov}, A.K.},
  \bibinfo{year}{2014}.
\newblock \bibinfo{title}{{RATAN-600 multi-frequency data for the BL Lacertae
  objects}}.
\newblock \bibinfo{journal}{\aap} \bibinfo{volume}{572}, \bibinfo{pages}{A59}.
\newblock \DOIprefix\doi{10.1051/0004-6361/201424437},
  \href{http://arxiv.org/abs/1410.2835}{{\tt arXiv:1410.2835}}.
\bibitem[{{Ott} et~al.(1994){Ott}, {Witzel}, {Quirrenbach}, {Krichbaum},
  {Standke}, {Schalinski} and {Hummel}}]{1994A&A...284..331O}
\bibinfo{author}{{Ott}, M.}, \bibinfo{author}{{Witzel}, A.},
  \bibinfo{author}{{Quirrenbach}, A.}, \bibinfo{author}{{Krichbaum}, T.P.},
  \bibinfo{author}{{Standke}, K.J.}, \bibinfo{author}{{Schalinski}, C.J.},
  \bibinfo{author}{{Hummel}, C.A.}, \bibinfo{year}{1994}.
\newblock \bibinfo{title}{{An updated list of radio flux density calibrators.}}
\newblock \bibinfo{journal}{\aap} \bibinfo{volume}{284},
  \bibinfo{pages}{331--339}.
\bibitem[{{Owen} et~al.(1989){Owen}, {Hardee} and {Cornwell}}]{Owen1989}
\bibinfo{author}{{Owen}, F.N.}, \bibinfo{author}{{Hardee}, P.E.},
  \bibinfo{author}{{Cornwell}, T.J.}, \bibinfo{year}{1989}.
\newblock \bibinfo{title}{{High-Resolution, High Dynamic Range VLA Images of
  the M87 Jet at 2 Centimeters}}.
\newblock \bibinfo{journal}{\apj} \bibinfo{volume}{340}, \bibinfo{pages}{698}.
\newblock \DOIprefix\doi{10.1086/167430}.
\bibitem[{{Paiano} et~al.(2017){Paiano}, {Falomo}, {Landoni}, {Treves} and
  {Scarpa}}]{paiano17}
\bibinfo{author}{{Paiano}, S.}, \bibinfo{author}{{Falomo}, R.},
  \bibinfo{author}{{Landoni}, M.}, \bibinfo{author}{{Treves}, A.},
  \bibinfo{author}{{Scarpa}, R.}, \bibinfo{year}{2017}.
\newblock \bibinfo{title}{{An optical view of extragalactic gamma-ray
  emitters}}.
\newblock \bibinfo{journal}{Frontiers in Astronomy and Space Sciences}
  \bibinfo{volume}{4}, \bibinfo{pages}{45}.
\newblock \DOIprefix\doi{10.3389/fspas.2017.00045},
  \href{http://arxiv.org/abs/1711.00325}{{\tt arXiv:1711.00325}}.
\bibitem[{{Pearson}(1896)}]{pearson}
\bibinfo{author}{{Pearson}, K.}, \bibinfo{year}{1896}.
\newblock \bibinfo{title}{{Mathematical Contributions to the Theory of
  Evolution. III. Regression, Heredity, and Panmixia}}.
\newblock \bibinfo{journal}{Philosophical Transactions of the Royal Society of
  London Series A} \bibinfo{volume}{187}, \bibinfo{pages}{253--318}.
\newblock \DOIprefix\doi{10.1098/rsta.1896.0007}.
\bibitem[{{Pian} et~al.(1998){Pian}, {Vacanti}, {Tagliaferri}, {Ghisellini},
  {Maraschi}, {Treves}, {Urry}, {Fiore}, {Giommi} and {Palazzi}}]{Pian98}
\bibinfo{author}{{Pian}, E.}, \bibinfo{author}{{Vacanti}, G.},
  \bibinfo{author}{{Tagliaferri}, G.}, \bibinfo{author}{{Ghisellini}, G.},
  \bibinfo{author}{{Maraschi}, L.}, \bibinfo{author}{{Treves}, A.},
  \bibinfo{author}{{Urry}, C.M.}, \bibinfo{author}{{Fiore}, F.},
  \bibinfo{author}{{Giommi}, P.}, \bibinfo{author}{{Palazzi}, E.},
  \bibinfo{year}{1998}.
\newblock \bibinfo{title}{{BeppoSAX Observations of Unprecedented Synchrotron
  Activity in the BL Lacertae Object Markarian 501}}.
\newblock \bibinfo{journal}{\apj} \bibinfo{volume}{492},
  \bibinfo{pages}{L17--L20}.
\newblock \DOIprefix\doi{10.1086/311083},
  \href{http://arxiv.org/abs/astro-ph/9710331}{{\tt arXiv:astro-ph/9710331}}.
\bibitem[{{Sambruna} et~al.(2001){Sambruna}, {Urry}, {Tavecchio}, {Maraschi},
  {Scarpa}, {Chartas} and {Muxlow}}]{Chandra_3C273}
\bibinfo{author}{{Sambruna}, R.M.}, \bibinfo{author}{{Urry}, C.M.},
  \bibinfo{author}{{Tavecchio}, F.}, \bibinfo{author}{{Maraschi}, L.},
  \bibinfo{author}{{Scarpa}, R.}, \bibinfo{author}{{Chartas}, G.},
  \bibinfo{author}{{Muxlow}, T.}, \bibinfo{year}{2001}.
\newblock \bibinfo{title}{{Chandra Observations of the X-Ray Jet of 3C 273}}.
\newblock \bibinfo{journal}{\apj} \bibinfo{volume}{549},
  \bibinfo{pages}{L161--L165}.
\newblock \DOIprefix\doi{10.1086/319157},
  \href{http://arxiv.org/abs/astro-ph/0101299}{{\tt arXiv:astro-ph/0101299}}.
\bibitem[{{Savolainen} et~al.(2010){Savolainen}, {Homan}, {Hovatta}, {Kadler},
  {Kovalev}, {Lister}, {Ros} and {Zensus}}]{Savolainen10}
\bibinfo{author}{{Savolainen}, T.}, \bibinfo{author}{{Homan}, D.C.},
  \bibinfo{author}{{Hovatta}, T.}, \bibinfo{author}{{Kadler}, M.},
  \bibinfo{author}{{Kovalev}, Y.Y.}, \bibinfo{author}{{Lister}, M.L.},
  \bibinfo{author}{{Ros}, E.}, \bibinfo{author}{{Zensus}, J.A.},
  \bibinfo{year}{2010}.
\newblock \bibinfo{title}{{Relativistic beaming and gamma-ray brightness of
  blazars}}.
\newblock \bibinfo{journal}{\aap} \bibinfo{volume}{512}, \bibinfo{pages}{A24}.
\newblock \DOIprefix\doi{10.1051/0004-6361/200913740},
  \href{http://arxiv.org/abs/0911.4924}{{\tt arXiv:0911.4924}}.
\bibitem[{{Scargle} et~al.(2013){Scargle}, {Norris}, {Jackson} and
  {Chiang}}]{bayes}
\bibinfo{author}{{Scargle}, J.D.}, \bibinfo{author}{{Norris}, J.P.},
  \bibinfo{author}{{Jackson}, B.}, \bibinfo{author}{{Chiang}, J.},
  \bibinfo{year}{2013}.
\newblock \bibinfo{title}{{Studies in Astronomical Time Series Analysis. VI.
  Bayesian Block Representations}}.
\newblock \bibinfo{journal}{\apj} \bibinfo{volume}{764}, \bibinfo{pages}{167}.
\newblock \DOIprefix\doi{10.1088/0004-637X/764/2/167},
  \href{http://arxiv.org/abs/1207.5578}{{\tt arXiv:1207.5578}}.
\bibitem[{{Schlafly} and {Finkbeiner}(2011)}]{S_F11}
\bibinfo{author}{{Schlafly}, E.F.}, \bibinfo{author}{{Finkbeiner}, D.P.},
  \bibinfo{year}{2011}.
\newblock \bibinfo{title}{{Measuring Reddening with Sloan Digital Sky Survey
  Stellar Spectra and Recalibrating SFD}}.
\newblock \bibinfo{journal}{\apj} \bibinfo{volume}{737}, \bibinfo{pages}{103}.
\newblock \DOIprefix\doi{10.1088/0004-637X/737/2/103},
  \href{http://arxiv.org/abs/1012.4804}{{\tt arXiv:1012.4804}}.
\bibitem[{{Sesar} et~al.(2007){Sesar}, {Ivezi{\'c}}, {Lupton}, {Juri{\'c}},
  {Gunn}, {Knapp}, {DeLee}, {Smith}, {Miknaitis}, {Lin}, {Tucker}, {Doi},
  {Tanaka}, {Fukugita}, {Holtzman}, {Kent}, {Yanny}, {Schlegel}, {Finkbeiner},
  {Padmanabhan}, {Rockosi}, {Bond}, {Lee}, {Stoughton}, {Jester}, {Harris},
  {Harding}, {Brinkmann}, {Schneider}, {York}, {Richmond} and {Vanden
  Berk}}]{sesar07}
\bibinfo{author}{{Sesar}, B.}, \bibinfo{author}{{Ivezi{\'c}}, {\v{Z}}.},
  \bibinfo{author}{{Lupton}, R.H.}, \bibinfo{author}{{Juri{\'c}}, M.},
  \bibinfo{author}{{Gunn}, J.E.}, \bibinfo{author}{{Knapp}, G.R.},
  \bibinfo{author}{{DeLee}, N.}, \bibinfo{author}{{Smith}, J.A.},
  \bibinfo{author}{{Miknaitis}, G.}, \bibinfo{author}{{Lin}, H.},
  \bibinfo{author}{{Tucker}, D.}, \bibinfo{author}{{Doi}, M.},
  \bibinfo{author}{{Tanaka}, M.}, \bibinfo{author}{{Fukugita}, M.},
  \bibinfo{author}{{Holtzman}, J.}, \bibinfo{author}{{Kent}, S.},
  \bibinfo{author}{{Yanny}, B.}, \bibinfo{author}{{Schlegel}, D.},
  \bibinfo{author}{{Finkbeiner}, D.}, \bibinfo{author}{{Padmanabhan}, N.},
  \bibinfo{author}{{Rockosi}, C.M.}, \bibinfo{author}{{Bond}, N.},
  \bibinfo{author}{{Lee}, B.}, \bibinfo{author}{{Stoughton}, C.},
  \bibinfo{author}{{Jester}, S.}, \bibinfo{author}{{Harris}, H.},
  \bibinfo{author}{{Harding}, P.}, \bibinfo{author}{{Brinkmann}, J.},
  \bibinfo{author}{{Schneider}, D.P.}, \bibinfo{author}{{York}, D.},
  \bibinfo{author}{{Richmond}, M.W.}, \bibinfo{author}{{Vanden Berk}, D.},
  \bibinfo{year}{2007}.
\newblock \bibinfo{title}{{Exploring the Variable Sky with the Sloan Digital
  Sky Survey}}.
\newblock \bibinfo{journal}{\aj} \bibinfo{volume}{134},
  \bibinfo{pages}{2236--2251}.
\newblock \DOIprefix\doi{10.1086/521819},
  \href{http://arxiv.org/abs/0704.0655}{{\tt arXiv:0704.0655}}.
\bibitem[{{Shappee} et~al.(2014){Shappee}, {Prieto}, {Grupe}, {Kochanek},
  {Stanek}, {De Rosa}, {Mathur}, {Zu}, {Peterson} and {Pogge}}]{Shappe14}
\bibinfo{author}{{Shappee}, B.J.}, \bibinfo{author}{{Prieto}, J.L.},
  \bibinfo{author}{{Grupe}, D.}, \bibinfo{author}{{Kochanek}, C.S.},
  \bibinfo{author}{{Stanek}, K.Z.}, \bibinfo{author}{{De Rosa}, G.},
  \bibinfo{author}{{Mathur}, S.}, \bibinfo{author}{{Zu}, Y.},
  \bibinfo{author}{{Peterson}, B.M.}, \bibinfo{author}{{Pogge}, R.W.},
  \bibinfo{year}{2014}.
\newblock \bibinfo{title}{{The Man behind the Curtain: X-Rays Drive the UV
  through NIR Variability in the 2013 Active Galactic Nucleus Outburst in NGC
  2617}}.
\newblock \bibinfo{journal}{\apj} \bibinfo{volume}{788}, \bibinfo{pages}{48}.
\newblock \DOIprefix\doi{10.1088/0004-637X/788/1/48},
  \href{http://arxiv.org/abs/1310.2241}{{\tt arXiv:1310.2241}}.
\bibitem[{{Tabara} and {Inoue}(1980)}]{1980A&AS...39..379T}
\bibinfo{author}{{Tabara}, H.}, \bibinfo{author}{{Inoue}, M.},
  \bibinfo{year}{1980}.
\newblock \bibinfo{title}{{A catalogue of linear polarization of radio
  sources.}}
\newblock \bibinfo{journal}{\aaps} \bibinfo{volume}{39},
  \bibinfo{pages}{379--393}.
\bibitem[{{Tramacere} et~al.(2007){Tramacere}, {Massaro} and
  {Cavaliere}}]{2007A&A...466..521T}
\bibinfo{author}{{Tramacere}, A.}, \bibinfo{author}{{Massaro}, F.},
  \bibinfo{author}{{Cavaliere}, A.}, \bibinfo{year}{2007}.
\newblock \bibinfo{title}{{Signatures of synchrotron emission and of electron
  acceleration in the X-ray spectra of Mrk 421}}.
\newblock \bibinfo{journal}{\aap} \bibinfo{volume}{466},
  \bibinfo{pages}{521--529}.
\newblock \DOIprefix\doi{10.1051/0004-6361:20066723},
  \href{http://arxiv.org/abs/astro-ph/0702151}{{\tt arXiv:astro-ph/0702151}}.
\bibitem[{{Udovitskiy} et~al.(2016){Udovitskiy}, {Sotnikova}, {Mingaliev},
  {Tsybulev}, {Zhekanis} and {Nizhelskij}}]{2016AstBu..71..496U}
\bibinfo{author}{{Udovitskiy}, R.Y.}, \bibinfo{author}{{Sotnikova}, Y.V.},
  \bibinfo{author}{{Mingaliev}, M.G.}, \bibinfo{author}{{Tsybulev}, P.G.},
  \bibinfo{author}{{Zhekanis}, G.V.}, \bibinfo{author}{{Nizhelskij}, N.A.},
  \bibinfo{year}{2016}.
\newblock \bibinfo{title}{{Automated system for reduction of observational data
  on RATAN-600 radio telescope}}.
\newblock \bibinfo{journal}{Astrophysical Bulletin} \bibinfo{volume}{71},
  \bibinfo{pages}{496--505}.
\newblock \DOIprefix\doi{10.1134/S1990341316040131}.
\bibitem[{{Urry} and {Padovani}(1995)}]{Urry95}
\bibinfo{author}{{Urry}, C.M.}, \bibinfo{author}{{Padovani}, P.},
  \bibinfo{year}{1995}.
\newblock \bibinfo{title}{{Unified Schemes for Radio-Loud Active Galactic
  Nuclei}}.
\newblock \bibinfo{journal}{\pasp} \bibinfo{volume}{107}, \bibinfo{pages}{803}.
\newblock \DOIprefix\doi{10.1086/133630},
  \href{http://arxiv.org/abs/astro-ph/9506063}{{\tt arXiv:astro-ph/9506063}}.
\bibitem[{{Verkhodanov}(1997)}]{1997ASPC..125...46V}
\bibinfo{author}{{Verkhodanov}, O.V.}, \bibinfo{year}{1997}.
\newblock \bibinfo{title}{{Multiwave Continuum Data Reduction at RATAN-600}},
  in: \bibinfo{editor}{{Hunt}, G.}, \bibinfo{editor}{{Payne}, H.} (Eds.),
  \bibinfo{booktitle}{Astronomical Data Analysis Software and Systems VI},
  p.~\bibinfo{pages}{46}.
\bibitem[{{Virtanen} et~al.(2019){Virtanen}, {Gommers}, {Oliphant},
  {Haberland}, {Reddy}, {Cournapeau}, {Burovski}, {Peterson}, {Weckesser},
  {Bright}, {van der Walt}, {Brett}, {Wilson}, {Jarrod Millman}, {Mayorov},
  {Nelson}, {Jones}, {Kern}, {Larson}, {Carey}, {Polat}, {Feng}, {Moore}, {Vand
  erPlas}, {Laxalde}, {Perktold}, {Cimrman}, {Henriksen}, {Quintero}, {Harris},
  {Archibald}, {Ribeiro}, {Pedregosa}, {van Mulbregt} and
  {Contributors}}]{scipy}
\bibinfo{author}{{Virtanen}, P.}, \bibinfo{author}{{Gommers}, R.},
  \bibinfo{author}{{Oliphant}, T.E.}, \bibinfo{author}{{Haberland}, M.},
  \bibinfo{author}{{Reddy}, T.}, \bibinfo{author}{{Cournapeau}, D.},
  \bibinfo{author}{{Burovski}, E.}, \bibinfo{author}{{Peterson}, P.},
  \bibinfo{author}{{Weckesser}, W.}, \bibinfo{author}{{Bright}, J.},
  \bibinfo{author}{{van der Walt}, S.J.}, \bibinfo{author}{{Brett}, M.},
  \bibinfo{author}{{Wilson}, J.}, \bibinfo{author}{{Jarrod Millman}, K.},
  \bibinfo{author}{{Mayorov}, N.}, \bibinfo{author}{{Nelson}, A.R.J.},
  \bibinfo{author}{{Jones}, E.}, \bibinfo{author}{{Kern}, R.},
  \bibinfo{author}{{Larson}, E.}, \bibinfo{author}{{Carey}, C.},
  \bibinfo{author}{{Polat}, {\.I}.}, \bibinfo{author}{{Feng}, Y.},
  \bibinfo{author}{{Moore}, E.W.}, \bibinfo{author}{{Vand erPlas}, J.},
  \bibinfo{author}{{Laxalde}, D.}, \bibinfo{author}{{Perktold}, J.},
  \bibinfo{author}{{Cimrman}, R.}, \bibinfo{author}{{Henriksen}, I.},
  \bibinfo{author}{{Quintero}, E.A.}, \bibinfo{author}{{Harris}, C.R.},
  \bibinfo{author}{{Archibald}, A.M.}, \bibinfo{author}{{Ribeiro}, A.H.},
  \bibinfo{author}{{Pedregosa}, F.}, \bibinfo{author}{{van Mulbregt}, P.},
  \bibinfo{author}{{Contributors}, S...}, \bibinfo{year}{2019}.
\newblock \bibinfo{title}{{SciPy 1.0--Fundamental Algorithms for Scientific
  Computing in Python}}.
\newblock \bibinfo{journal}{arXiv e-prints} ,
  \bibinfo{pages}{arXiv:1907.10121}\href{http://arxiv.org/abs/1907.10121}{{\tt
  arXiv:1907.10121}}.
\bibitem[{{Wang} and {Zhou}(2009)}]{wang09}
\bibinfo{author}{{Wang}, C.C.}, \bibinfo{author}{{Zhou}, H.Y.},
  \bibinfo{year}{2009}.
\newblock \bibinfo{title}{{Determination of the intrinsic velocity field in the
  M87 jet}}.
\newblock \bibinfo{journal}{\mnras} \bibinfo{volume}{395},
  \bibinfo{pages}{301--310}.
\newblock \DOIprefix\doi{10.1111/j.1365-2966.2009.14463.x},
  \href{http://arxiv.org/abs/0904.1857}{{\tt arXiv:0904.1857}}.
\bibitem[{{Weymann} et~al.(1991){Weymann}, {Morris}, {Foltz} and
  {Hewett}}]{weymann91}
\bibinfo{author}{{Weymann}, R.J.}, \bibinfo{author}{{Morris}, S.L.},
  \bibinfo{author}{{Foltz}, C.B.}, \bibinfo{author}{{Hewett}, P.C.},
  \bibinfo{year}{1991}.
\newblock \bibinfo{title}{{Comparisons of the Emission-Line and Continuum
  Properties of Broad Absorption Line and Normal Quasi-stellar Objects}}.
\newblock \bibinfo{journal}{\apj} \bibinfo{volume}{373}, \bibinfo{pages}{23}.
\newblock \DOIprefix\doi{10.1086/170020}.
\bibitem[{{Wilson} and {Yang}(2002)}]{Chandra_M87}
\bibinfo{author}{{Wilson}, A.S.}, \bibinfo{author}{{Yang}, Y.},
  \bibinfo{year}{2002}.
\newblock \bibinfo{title}{{Chandra X-Ray Imaging and Spectroscopy of the M87
  Jet and Nucleus}}.
\newblock \bibinfo{journal}{\apj} \bibinfo{volume}{568},
  \bibinfo{pages}{133--140}.
\newblock \DOIprefix\doi{10.1086/338887},
  \href{http://arxiv.org/abs/astro-ph/0112097}{{\tt arXiv:astro-ph/0112097}}.
\bibitem[{{Wright} et~al.(2010){Wright}, {Eisenhardt}, {Mainzer}, {Ressler},
  {Cutri}, {Jarrett}, {Kirkpatrick}, {Padgett}, {McMillan} and
  {Skrutskie}}]{Wright10}
\bibinfo{author}{{Wright}, E.L.}, \bibinfo{author}{{Eisenhardt}, P.R.M.},
  \bibinfo{author}{{Mainzer}, A.K.}, \bibinfo{author}{{Ressler}, M.E.},
  \bibinfo{author}{{Cutri}, R.M.}, \bibinfo{author}{{Jarrett}, T.},
  \bibinfo{author}{{Kirkpatrick}, J.D.}, \bibinfo{author}{{Padgett}, D.},
  \bibinfo{author}{{McMillan}, R.S.}, \bibinfo{author}{{Skrutskie}, M.},
  \bibinfo{year}{2010}.
\newblock \bibinfo{title}{{The Wide-field Infrared Survey Explorer (WISE):
  Mission Description and Initial On-orbit Performance}}.
\newblock \bibinfo{journal}{\aj} \bibinfo{volume}{140},
  \bibinfo{pages}{1868--1881}.
\newblock \DOIprefix\doi{10.1088/0004-6256/140/6/1868},
  \href{http://arxiv.org/abs/1008.0031}{{\tt arXiv:1008.0031}}.
\bibitem[{{Zdziarski} et~al.(1995){Zdziarski}, {Johnson}, {Done}, {Smith} and
  {McNaron-Brown}}]{Zdziarski95}
\bibinfo{author}{{Zdziarski}, A.A.}, \bibinfo{author}{{Johnson}, W.N.},
  \bibinfo{author}{{Done}, C.}, \bibinfo{author}{{Smith}, D.},
  \bibinfo{author}{{McNaron-Brown}, K.}, \bibinfo{year}{1995}.
\newblock \bibinfo{title}{{The Average X-Ray/Gamma-Ray Spectra of Seyfert
  Galaxies from GINGA and OSSE and the Origin of the Cosmic X-Ray Background}}.
\newblock \bibinfo{journal}{\apj} \bibinfo{volume}{438}, \bibinfo{pages}{L63}.
\newblock \DOIprefix\doi{10.1086/187716}.
\bibitem[{{Zhang} et~al.(2011){Zhang}, {Wang}, {Zhou}, {Wang} and
  {Jiang}}]{zhang11}
\bibinfo{author}{{Zhang}, S.H.}, \bibinfo{author}{{Wang}, H.Y.},
  \bibinfo{author}{{Zhou}, H.Y.}, \bibinfo{author}{{Wang}, T.G.},
  \bibinfo{author}{{Jiang}, P.}, \bibinfo{year}{2011}.
\newblock \bibinfo{title}{{Discovery of a variable broad absorption line in the
  BL Lac object PKS B0138-097}}.
\newblock \bibinfo{journal}{Research in Astronomy and Astrophysics}
  \bibinfo{volume}{11}, \bibinfo{pages}{1163--1170}.
\newblock \DOIprefix\doi{10.1088/1674-4527/11/10/005},
  \href{http://arxiv.org/abs/1106.1587}{{\tt arXiv:1106.1587}}.

\end{thebibliography}

\end{document}